\documentclass[12pt,man, floatsintext]{apa7}
\usepackage[american]{babel}

\usepackage[T1]{fontenc} 
\usepackage[latin1]{inputenc}

\usepackage{blindtext}

\usepackage{graphicx}
\usepackage{natbib}
\usepackage{amsmath}

\usepackage{multirow}
\usepackage{array}

\usepackage{subfiles}
\usepackage{subcaption}
\usepackage{wrapfig}

\usepackage{booktabs,makecell}

\usepackage{tikz}
\usetikzlibrary{shapes, arrows, fit, positioning}
\usetikzlibrary{shapes.geometric}
\usepackage{pgf} 

\usepackage[normalem]{ulem}

\usepackage{setspace}

\usepackage{fancyhdr}
\usepackage{blkarray}
\usepackage{listings}
\usepackage{adjustbox}

\usepackage{url}
\usepackage{hyperref}

\begin{document}

\title{Data Quality in Crowdsourcing and Spamming Behavior Detection} 
\shorttitle{Data Quality} 


\authorsnames[1, 2, 3, 1]{Yang Ba,  Michelle V. Mancenido, Erin K. Chiou,  Rong Pan}
\authorsaffiliations{
{School of Computing and Augmented Intelligence, Arizona State University, AZ, USA}, 
{School of Mathematical and Natural Sciences, Arizona State University, AZ, USA}, 
{Human Systems Engineering, Arizona State University, AZ, USA}
}

\authornote{
   \addORCIDlink{Yang Ba}{0000-0000-0000-0000}

  Correspondence concerning this article should be addressed to Yang Ba, 
Ira A. Fulton Schools of Engineering, School of Computing and Augmented Intelligence, Data Science, Analytics and Engineering, Arizona State University, Suite 342AE, 3rd floor 699 S. Mill Avenue
Tempe, AZ 85281, E-mail: yangba@asu.edu, 623-272-3499}







\abstract{As crowdsourcing emerges as an efficient and cost-effective method for obtaining labels for machine learning datasets, it is important to assess the quality of crowd-provided data, so as to improve analysis performance and reduce biases in subsequent machine learning tasks. Given the lack of ground truth in most cases of crowdsourcing, we refer to data quality as annotators' consistency and credibility. Unlike the simple scenarios where Kappa coefficient and intraclass correlation coefficient usually can apply, online crowdsourcing requires dealing with more complex situations. We introduce a systematic method for evaluating data quality and detecting spamming threats via variance decomposition, and we classify spammers into three categories based on their different behavioral patterns. A spammer index is proposed to assess entire data consistency, and two metrics are developed to measure crowd workers' credibility by utilizing the Markov chain and generalized random effects models. Furthermore, we showcase the practicality of our techniques and their advantages by applying them on a face verification task with both simulation and real-world data collected from two crowdsourcing platforms.
}

\keywords{crowdsourcing platform, data quality, spamming behaviors, metrics, generalized random effects models, statistical hypothesis testing} 

\newcommand{\mickey}[1]{\textcolor{red}{#1}}

\maketitle

\section{Introduction}

The emergence of crowdsourcing as an online platform for data collection has raised concerns about data integrity due to its widespread popularity and low barriers to entry. Crowdsourcing involves engaging web-based (or crowdsourcing) workers to voluntarily undertake a range of tasks, from simple surveys to complex digital experiments, leveraging collective human intelligence to test research hypotheses or to perform manual labeling \citep{estelles2015crowdsourcing}.
This approach has gained considerable traction in the social and behavioral sciences as a cost-effective and expeditious method for conducting experimental research \citep{meyer2016net}. 
For example, a 2016 study reported a 10--30\% increase in the use of Amazon's Mechanical Turk platform among the top 3 psychology journals between 2012 to 2015 \citep{zhou2016pitfall}. Recently, niche platforms such as Prolific.co have risen in popularity among academic researchers in data-intensive domains, advertising tailored services for filtering the participant pool to increase the chances of credible and reliable responses \citep{peer2022data}.

However, crowdsourced data, when compared to data collected from domain experts and proctored experiments, often display significant variability in quality \citep{hsueh2009data}. This problem is well-documented in the computer sciences, where crowdsourcing is typically used to obtain annotations or labels, such as images and text, to facilitate the creation of machine learning (ML) datasets. Responses from multiple crowd workers are usually distilled into a ground truth using aggregation methods such as majority voting \citep{tao2018domain}. These labeled datasets are then fed into ML algorithms, enabling them to discern patterns and make predictions. For instance, state-of-the-art models are able to identify a pedestrian in an image cluttered with vehicles and buildings \citep{gm2021urban}, or subjectively classify sentences with racial biases \citep{waseem2016you}. Data variability can lead to a decline in the performance of ML models trained on such data \citep{sheng2019machine, li2016noise}. For online experiments, the potential variability in data quality can obscure, amplify, or confound the true effects of the factors under investigation, potentially threatening the validity and generalizability of scientific results \citep{steger2018meta}.

\begin{figure}[ht]
\caption{Experimental Diagrams}
\centering
\begin{subfigure}{\textwidth}
  \centering
  \includegraphics[width=0.48\linewidth]{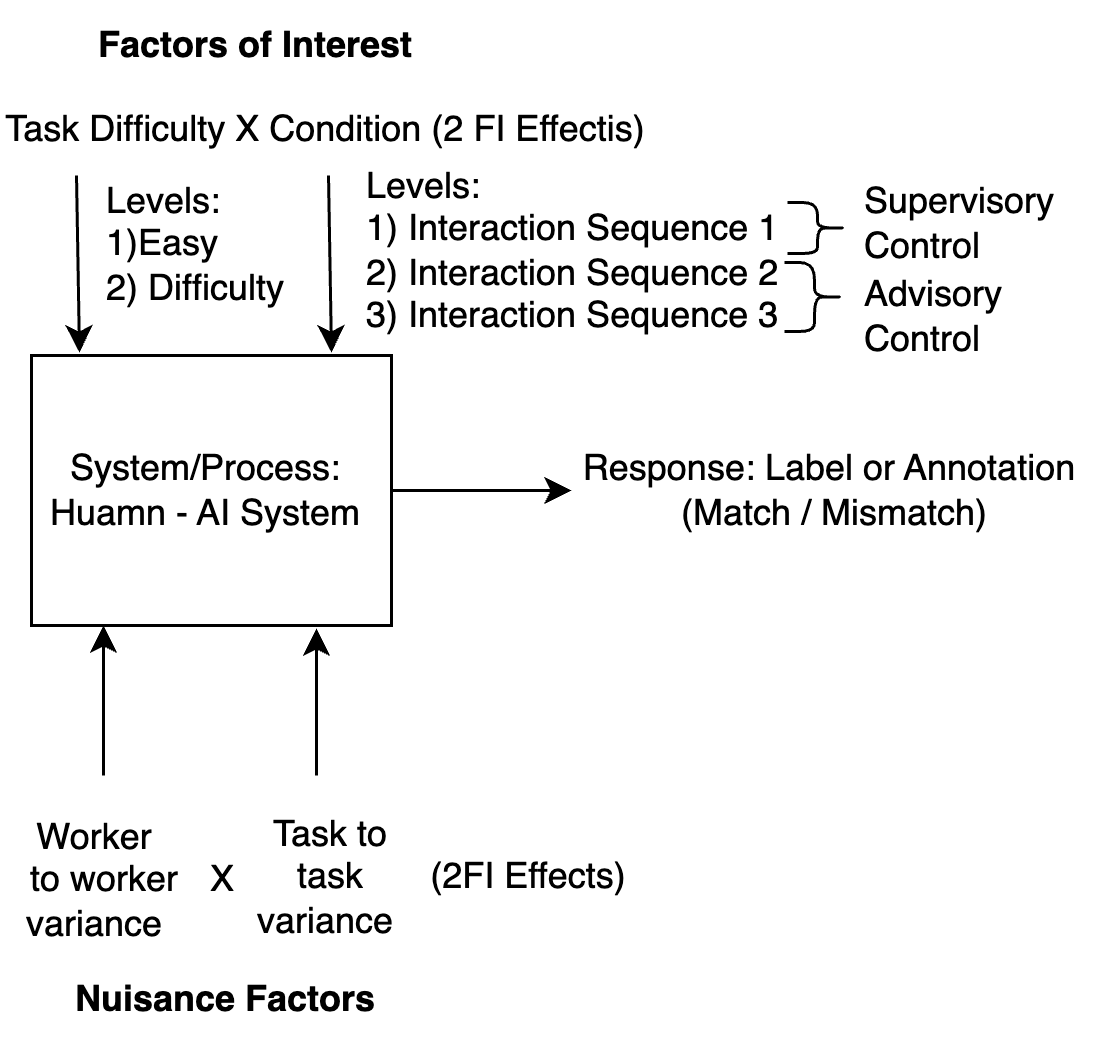}
  \caption{Experimental Diagram for the MTurk Dataset. Two-factor interactions (2FI) are the highest-order effects (for both random and fixed effects) of interest. For the Prolific dataset, only two levels of "Condition" were of interest namely, Supervisory VS. Advisory control.}
  \label{fig:ED1}
\end{subfigure}\\ 

\begin{subfigure}{\textwidth}
  \centering
  \includegraphics[width=0.48\linewidth]{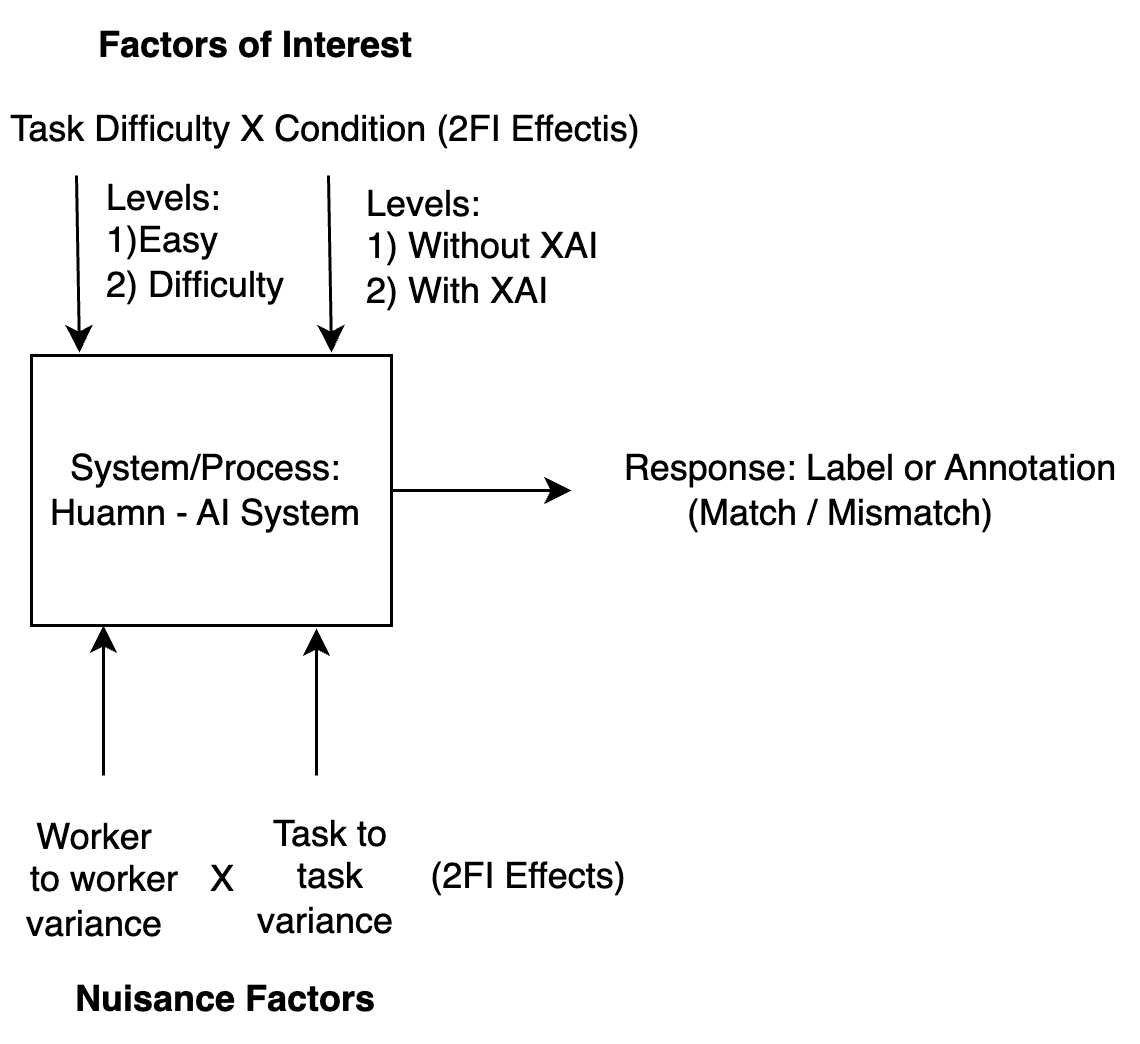}
  \caption{Data collected in-person (airport data). In this experiment, "Condition" refers to the type of AI-enabled face verification system deployed. The "without XAI" condition only shows model outputs to participants. The "with XAI" condition shows both model outputs and model confidence, alongside a heat map (saliency map) to highlight facial regions used by the AI Model for prediction.}
  \label{fig:ED2}
\end{subfigure}

\label{fig:ED}
\end{figure}

The motivating cases for this research are three experiments on AI-enabled systems conducted by the authors \citep{salehi2023evaluating}, wherein two were deployed in crowdsourcing platforms and the third conducted in-person at different airports as part of a funded project with the Department of Homeland Security.  
A web-based platform was developed to emulate an automated face recognition system (AFRS) with a human-in-the-loop 
\footnote{https://github.com/YangBa78/Facewise-Readit/tree/main/facewise}. 
Participants were tasked to perform identity verification of two photos, the expected response being a "match" if the photos were of the same person or a "mismatch" if they were of different identities. The experiment was primarily focused on characterizing the effects of the manipulated factors related to the AFRS ("condition") on annotation performance (see Figure \ref{fig:ED}). Further, the complexity of face matching tasks was empirically manipulated to be easy or difficult. These design variables are called "fixed effects" or factors of interest in experimental design theory \citep{montgomery2017design}. Further, following principled guidelines for designing statistically valid experiments, study variables that were not directly linked to the experimental hypotheses but affect the annotations or responses were also identified. These variables, often called nuisance or random effects, include task-to-task differences within a difficulty level and variation among participant competencies. Human-technology interaction studies, such as the one just described, are commonly deployed in the internet using crowdsourcing platforms \citep{peng2013mapping}.

In these studies, fixed effects are the primary study parameters that are posited to impact the response or outcome. As secondary parameters, random effects are a result of nuisance variables that are known \emph{a priori} to have an impact on the outcome; however, typical strategies for alleviating their variational effects, such as randomization, may not be sufficient due to the unproctored and widely accessible nature of crowdsourced data collection. Thus, these nuisance variables are generally expected to contribute to the overall experimental error, which will affect the standard errors of fixed effect estimates and consequently, tests of significance.

While data quality in crowdsourcing platforms spans multi-faceted definitions such as accuracy \citep{hsueh2009data} or completeness \citep{van2010impact}, we limit our definition of data quality as the ability of annotators to provide \emph{consistent} and \emph{credible} responses over the course of data collection (for succinctness, we refer to this as \textit{worker reliability}). Consistency and credibility are constructs related to the capability of the measurement system to distinguish among intended variation (such as differences in control and treatment groups), variation due to random chance or experimental error, and measurement error.

\textit{Consistency}, particularly when there's no definitive ground truth or gold standard, reflects the degree of concordance among multiple annotators. Existing approaches for gauging consistency include the intra-class correlation coefficient (ICC) for numeric ratings \citep{mehta2018performance} and the Kappa coefficient for categorical annotations \citep{perez2016building, chmura2002kappa, junkes2015validity}. These metrics, derived from classical test theory, exhibit some limitations in their applicability to crowdsourced data similar to the motivating case. The ICC, which applies variance decomposition methods akin to gage repeatability and reproducibility studies in quality control \citep{montgomery2009statistical}, is designed for truly numeric data. For categorical ratings or annotations such as in identity verification, the Kappa coefficient is a more suitable metric for reporting consistency. However, it bears its own set of limitations; it has been shown, for example, to underestimate true inter-rater reliability in the presence of a dominant category (e.g., unbalanced number of matches and mismatches \citep{kappaharm, feinstein1990high, dettori2020kappa}).

The underlying issue with the Kappa statistic's reliance on the assumption of independent and randomly distributed responses becomes particularly apparent in crowdsourced environments. Such platforms are prone to attracting annotators who may exhibit systematic, non-credible behaviors. Notorious among these are "spammers" who might, for example, consistently select the first response category. Such behaviors, which are driven not by thoughtful engagement to the task but by the pursuit of rapid financial gain, are overlooked by the Kappa statistic. Prior work has shown its tendency to be overinflated in the presence of spammers \citep{hall2022quality}. The metric calculates adjusted pairwise agreements which could artificially be inflated by spamming behaviors that disproportionately favor one category, leading to a high observed agreement that does not reflect true concordance. Thus, in addition to consistency, we must also analyze indicators of \textit{credibility} of responses as a vital component of measuring the integrity of crowdsourced datasets by detecting potential spamming behaviors.

In this paper, we propose an integrated framework of metrics to retrospectively evaluate the \textit{consistency} and \textit{credibility} of crowdsourced data. These methods were particularly developed for online experiments in which preemptive strategies, such as attention checks, may not be sufficient in preventing non-credible responses due to the absence of study proctors. However, the methods proposed in this paper can also apply to manual annotation or labeling tasks for the purposes of generating ML training and testing sets. 

The framework is designed to: (1) quantify measurement errors resulting from task-to-task, annotator-to-annotator, and task-within-annotator variation by applying a variance decomposition approach called generalized linear random effects model (GLRM); (2) detect and classify potential spamming behavior using Markov chains, KL divergence, and deletion analysis. We validate the efficacy of the proposed metrics for these tasks through simulation experiments and empirical trials conducted on both online and in-person platforms. Further, we use our proposed approach to corroborate the findings of a prior study \citep{peer2022data} that compares worker reliability between two online platforms, namely, Amazon's Mechanical Turk (MTurk) and Prolific.co.

The proposed framework draws inspiration from classical test theory methodologies, such as item response theory \citep{demars2010item} and generalizability theory \citep{webb20064}, yet it offers distinct benefits over traditional metrics. Unlike the Intra-class Correlation (ICC) commonly used in IRT, our method is versatile, capable of handling a diverse class of measurement scales (including binary and nominal responses) that frequently occur in crowdsourced data. Moreover, our use of the GLRM approach facilitates a unified analysis, merging data quality assessment with the estimation of random effects within a single modeling construct. Finally, our method broadens existing approaches by identifying additional indicators of anomalous behaviors in spammers, such as repeated patterns and random guessing.

\section{Related Work}

A significant portion of research into data quality for crowdsourced tasks has focused on the evaluation of preemptive strategies. These strategies range from employing bot-screening techniques \citep{rodriguez2023creating}, assessing worker competency through their track record \citep{gadiraju2017using}, mitigating effects of worker fatigue \citep{zhang2018understanding}, enhancing task engagement \citep{liang2018intrinsic}, to optimizing task design \citep{ceschia2022task}. Our proposal, however, takes a retrospective stance, focusing on the post-hoc analysis of data from completed experiments to identify patterns of spamming and subtle behavioral anomalies potentially overlooked by preventive strategies. 

Other retrospective approaches in the literature focus on arriving at a consensus for annotations, while taking into account the presence of noise and other exogenous variables that affect data quality. Central to this endeavor are methods such as the Dawid-Skene (DS) algorithm, which applies expectation-maximization (EM) to estimate true labels by modeling the error rates of annotators \citep{dawid1979maximum}. These methodologies, rooted in psychometrics and reliability theory, typically use metrics like inter-annotator agreement and intra-class correlation to measure \emph{consistency}. Thus, in the literature, consistency is often associated with \textit{how often annotators agree with themselves and among each other}. Yet, they may not fully account for deliberate low-effort contributions or patterned behavior that undermine data integrity.

Recent advances have introduced adaptive algorithms that iteratively adjust to the data submission patterns of contributors, refining quality estimates over time \citep{ipeirotis2014repeated}. This reflects a shift toward dynamic assessment models that consider the temporal and behavioral aspects of annotation, which are critical in understanding and leveraging crowd-worker diversity. In practice, however, researchers are faced with budget constraints and it is typical to run online experiments with only one set of workers. Thus, while dynamic assessment models are useful in applications where tasks are laborious and repetitive, such as in image labeling, they have limited usefulness for experiments wherein cost constraints prevent researchers from separately evaluating potential participants. 

In the following subsections, we examine existing metrics and methods for measuring the \textit{consistency} and \textit{credibility} of crowdsourced data. This discussion aims to set the stage for our contribution, which proposes novel metrics and modeling strategies to enhance the detection and treatment of low-quality data in crowdsourced environments.

\subsection{Measures for Consistency}

Classical Test Theory (CTT), first developed in the early 20th century, has been a cornerstone in behavioral and social science research for evaluating the reliability and validity of measurement instruments through its unique focus on observed test scores and error components~\citep{traub1997classical, webb20064}. In CTT, the variance of observed scores ($V(X)$) is decomposed into the variance of the true score ($V(T)$) plus the variance due to measurement error ($V(E)$) i.e., $ Var(X) = Var(T) + Var(E)$. The intra-class correlation coefficient (ICC), a common metric used to evaluate the consistency of measurement systems, which include annotators or operators, is calculated as: 

\begin{equation} 
\label{eq1}
ICC = \frac{\sigma_B^2} {\sigma_{total}^2} = \frac{\sigma_B^2} {\sigma_{B}^2 + \sigma_{\epsilon}^2}
\end{equation}

The idea behind the ICC is to quantify how strongly objects in the same class are grouped together in a measurement \citep{riezler2022validity}. In Equation \ref{eq1}, the ICC denotes the ratio of the variance between objects of interest $\sigma_{B}^2$ to the total variance $\sigma_{total}^2$, which can also be interpreted as the amount of variation coming from a source of interest. In the context of our motivating problem, the source of interest could be the variation coming from worker-to-worker differences in annotations. Thus, the ICC could be interpreted as the proportion of total variance attributed to inconsistencies in annotations among crowdsourcing workers. Some known limitations of the ICC are their appropriateness only for numeric-scale data \citep{mehta2018performance} and their weaknesses in small-sample and unbalanced data scenarios \citep{kappaharm}.

Another widely-used statistical measure for assessing the consistency of annotations or observed responses is the Kappa coefficient. In comparison to the ICC, Kappa is commonly applied to categorical data scales, such as binary or nominal. Its suitability for categorical data is inherent in its calculation: in lieu of comparing variances from sources of interest, Kappa considers chance agreement and adjusts its calculation for chance correction. Two of the more popular variants are Cohen's Kappa \citep{mchugh2012interrater}, which works best when there are two raters; and Fleiss Kappa \citep{falotico2015fleiss}, an extension of Cohen's for any number of raters. Cohen's Kappa is mathematically equivalent to ICC for binary data \citep{fleiss1973equivalence}.

Previous works have presented several weaknesses and limitations of the ICC and Kappa for evaluating crowdsourced data. In some cases, the Kappa score yields conflicting results, posting low values even in the presence of high agreement among annotators \citep{kappaharm}. Further, the Kappa score has been documented to be sensitive to the number of classes and raters, i.e., a lower number of annotators and classes have been found to be subject to higher chances of random agreement \citep{kappaharm}. Additionally, kappa scores have the same class skew issue as ICC, which could affect chance correction, because more errors are expected for the majority classes than the rare classes. In a more general examination of using ICC and Fleiss' Kappa for crowdsourced ratings, \cite{salminen2018inter} found that the nature of the task impacts agreement ratings, i.e., the more \textit{subjective} the task is, the lower the agreement. This suggests that agreement-based metrics are not necessarily an effective, sole measure of overall data integrity.

Finally, both ICC and Kappa scores face interpretation challenges. While some empirical guidelines exist on categorizing whether consistency is poor, fair, good, or excellent based on the calculated values of the metrics, there is no universal recommendation for determining the relationship strength between agreement and calculated scores. For example, guidelines in the original paper \citep{landis1977measurement} only focus on binary data, while many applications in crowdsourced environments yield nominal (e.g., class annotations) or ordinal (e.g., survey responses) data. 

\subsection{Measures for Credibility: Spammer Detection}

Compared to credible online workers or credible online study participants (henceforth "worker" or "workers" for simplicity), spammers are distinguished through displayed systematic behavioral patterns, such as random guessing or choosing a single, consistent answer regardless of task demands. Spammers can also be bots (e.g., software automation), commonly found on crowdsourced platforms. Other types of spamming behaviors are classified and categorized in \citep{gadiraju2015understanding}, which proposed a "maliciousness" index (MI) for classifying workers as untrustworthy or trustworthy based on the MI. MI is essentially a response acceptability measure that focuses on open-ended question responses. While \cite{gadiraju2015understanding} validated their work based on average task completion time, indicating that time would have a positive connection with spamming behaviors, the efficacy of time would be argued as a validation metric: participants can achieve good performance with relatively less time spent. Therefore, we only use average task completion time as an auxiliary factor for identifying spammers, serving as a part of assessing workers' credibility. 
Correspondingly, we validate our methods against accuracy on well-defined tasks, which provides a triangulating measure of effectiveness.

Among probabilistic methods, MACE (Multi-Annotator Competence Estimation) has been proposed as a graphical model for identifying spammers via unsupervised learning \citep{hovy2013learning}. MACE models and estimates the credibility of annotators and their spamming preferences as latent variables via a generative process, an approach widely used for linguistic data. In comparison to other methods, MACE provides a more accurate estimate of true labels to reduce the noise of data collected in crowdsourcing. However, while MACE considers only one type of spammer, the type that always selects a specific preferred answer (we call these primary-choice spammers), our approach can tackle a wide range of spamming behaviors that notably includes a more challenging to detect behavior, random guessing.

\section{Proposed Metrics}

In this section, we propose a set of metrics for measuring worker consistency and detecting threats to the credibility of crowdsourced data. Some of these metrics are variance ratio-based, similar to the ICC in which consistency is gauged by the ratio of variance pertaining to the variable of interest against the overall variance, while other metrics are models-based, leveraging log-likelihood derived from the identical model that assesses worker consistency and statistical significance (p-value) to indicate which responses are spammers. We then introduce an alternative approach to detect spamming behaviors, which involves classifying common spamming behaviors and exploring the correlation between the efficacy of consistency measurement and the detection of potential spamming threats. Our results indicate the importance of assessing data quality from two facets: \textit{consistency} and \textit{credibility}.

\subsection{Metric - Consistency}

Classic Test Theory (CTT) and the intra-class correlation coefficient (ICC) are metrics employing the variance ratio concept, wherein the variance of interest is divided by the total variance. We adopt this framework to assess the \emph{consistency} of crowdsourced data. In the context of crowdsourcing tasks, both tasks and participants are typically drawn randomly from their respective populations. Considering a simple scenario, a single crowd worker is assigned to handle multiple tasks, while each task is completed by multiple workers. Thus, the total variance contains the variance of workers, the variance of tasks, and the variance of interactions between them. We propose a variance ratio-based metric called \textit{Spammer Index} to measure the overall data quality.

\begin{equation}
\label{eq2}
\textrm{Spammer Index}  = \frac{\sigma^2_{workers}} {\sigma^2_{workers} + \sigma^2_{tasks} + \sigma^2_{workers: tasks}} 
\end{equation} \\

Compared to the intra-class correlation coefficient (ICC) in equation (1), there is no residual variance $\sigma^2_{\epsilon}$, since generalized linear models do not estimate error terms in the same way as linear regression models do for numerical data. \cite{browne2005variance} discusses methods for the estimation of the variance of error terms in the non-linear models, focusing primarily on logistic models, including simulation and the latent variable approach. The simulation method requires extra computations while the latent variable approach treats the binary response variable (0, 1) coming from an underlying continuous variable and applies the variance for the standard logistic distribution, 3.29. However, this variance is fixed and serves as a regularization term without a significant impact on the relative degree of the whole variance ratio. Therefore, for the sake of computing simplicity and broader applicability, we exclude error terms from \textit{Spammer Index}. We employ generalized linear random-effects models (GLRM) to capture the variance components since random effects are of interest. We exclude fixed predictors because they are not relevant variables that determine data quality in crowdsourcing as shown in Figure \ref{fig:ED}. Though we mainly discuss empirical experiments involving binary response variables, we can extend our analysis to multi-class cases through additional simulations, which demonstrate the broader applicability of our method.

For binary responses, the GLRM specification of random effects is given as below:
\begin{equation}
\label{eq3}
\begin{gathered}
        Y_{ij} \sim Bernoulli(p_{ij});  \\
        ln\left(\frac{p_{ij}}{1-p_{ij}}\right) = \beta_{0} + w_{i} + t_{j} + wt_{ij}
\end{gathered}
\end{equation}
where i is the index of a crowd worker (i = 1, ...,i); j is the index for a task (j =1, ..., j). Let $Y_{ij}$ equal to 1 for one of the binary options and 0 for the other and $p_{ij}$ is the probability of response of worker i for task j; $\beta_{0}$ is the intercept parameter. $w_{i}$ is the random effect of worker i and $w_{i} \sim N (0, \sigma^2_{workers})$; $t_{i}$ is the random effect of task j and $ t_{j} \sim N (0, \sigma^2_{tasks})$; $wt_{ij}$ is the random effect of worker i at task j and $wt_{ij} \sim N (0, \sigma^2_{workers: tasks})$; 

When crowd workers are more likely to agree with themselves or other workers, the variance among workers is small, and vice versa. Therefore, to have quality data in crowdsourcing, the variance of workers should account for a small part of the total variance, and the variance of tasks should dominate the total variance. On the other hand, when there is a higher variance among workers, it indicates less agreement among them and suggests the presence of less reliable crowd workers. Thus, a higher \emph{Spammer Index} informs that there is a higher likelihood of data contamination by spamming or other unreliable contributions from workers.

For ordinal classification, where response categories have a natural order (e.g., Likert scale responses include "Strongly Disagree", "Disagree", "Neutral", "Agree", "Strongly Agree"), the original Spammer Index formula (Equation \ref{eq2}) remains valid. Instead of modeling a single probability in binary responses, generalized linear random-effects model (GLRM) models cumulative probability for ordinal responses.
\[
\text{Logit}(P(y_{ij} \leq k)) = r_k - (w_i + t_j + w_{ti})
\]
The model estimates multiple threshold parameters $r_k$, which determine the cut-off points between categories. As a result, the random effects remain structured as follows: $w_{i} \sim N (0, \sigma^2_{workers})$, $ t_{j} \sim N (0, \sigma^2_{tasks})$, and $wt_{ij} \sim N (0, \sigma^2_{workers: tasks})$, introducing a single variance component for each random effect. Therefore, no modifications are required for the \textit{Spammer Index} equation.

In contrast, for nominal classification, where categories are unordered (e.g., annotating object types in images or identifying named entities in text annotations), the probability of each category is modeled independently, with each category (except the reference category) having its own separate random effect variance. To account for the unique category-level parameters and the lack of stochastic ordering, the \textit{Spammer Index} is modified as follows:

\begin{equation}
\label{eq4}
\textrm{Spammer Index}_{\textrm{nominal}} = \frac{\sum_{k} \sigma^2_{\textrm{workers}, k}}{\sum_{k} (\sigma^2_{\textrm{workers}, k} + \sigma^2_{\textrm{tasks}, k} + \sigma^2_{\textrm{workers:tasks}, k})}
\end{equation}
where $k$ represents each of the $K$ response categories. $\sigma^2_{\textrm{workers}, k}, \sigma^2_{\textrm{tasks}, k}, \sigma^2_{\textrm{workers: tasks}, k}$ captures variance among workers,  task-related variance and interaction variance between workers and tasks specifically for category $k$ respectively. By summing over all $K$ categories, the numerator reflects total worker variance, while the denominator accounts for overall variance contributions from workers, tasks, and worker-task interactions.

In more complicated online experiments, experiment designers may require crowd workers to annotate the same tasks multiple times by shuffling the order of tasks. They can obtain more precise responses by aggregating the same workers' annotations from different time stamps. In this scenario, the total variance of \emph{Spammer Index} includes $\sigma^2_{time}$, $\sigma^2_{workers: time}$ and $\sigma^2_{tasks: time}$.

\subsection{Metric - Credibility}
\subsubsection{Spamming Behaviors Classification}

As a metric to assess the level of agreement among workers in the context of crowdsourcing, the \textit{Spammer Index} may not encompass all information related to the \emph{credibility} of workers. Rather, the \textit{Spammer Index} is closely related to the distribution of worker responses and provides insights into the consistency or inconsistency of their responses. However, response patterns of spammers and unreliable workers vary. Our paper considers three typical spamming behaviors: \textit{Primary Choice}, \textit{Repeated Pattern}, and \textit{Random Guessing}, described in more detail below.

\begin{itemize}
  \item \textbf{Primary Choice:} The spammer has a strong preference to choose a specific response; i.e., they usually do not pay attention to the question instructions and always select the same answer. Their responses have a skewed distribution and a long run with the same answer(s) continuously, e.g., 1 0 0 0 0 0 0 0 0 0 0 ...
  \item \textbf{Repeated Pattern:} The spammer switches their answers and repeats this type of behavior frequently through all tasks, e.g.: 0 1 0 1 0 1 0 1 ... 
  \item \textbf{Random Guessing:} The spammer randomly picks their responses for each task. There is no obvious pattern for their responses. Their behavior patterns are very similar to those of credible workers if the tasks are also assigned randomly. 
\end{itemize}

It is generally assumed that spammers do not adhere to task instructions and aim to complete tasks quickly with minimal effort, often to maximize financial reward \citep{gadiraju2015understanding}. As their primary motivation is to maximize the earnings per time or effort unit, we can generally assume that they have no malicious intent to purposefully harm the data quality and they are unlikely to engage in more complicated spamming schemes. Thus, we focus on identifying the patterns associated with the aforementioned simple spamming behaviors. We also recognize that spamming-like behaviors may emerge midway through tasks due to boredom or fatigue. We investigate this case in Appendix C. While it is not a universal truth in every scenario, spammers often exhibit lower accuracy compared to credible workers, making accuracy a potential standard for validating our approaches.

\begin{figure}[ht]
\caption{\label{fig:pct} Plots of Spammer Index versus the intensity of spamming for different spamming behavior types}
\centering
\includegraphics[width=0.6\textwidth]{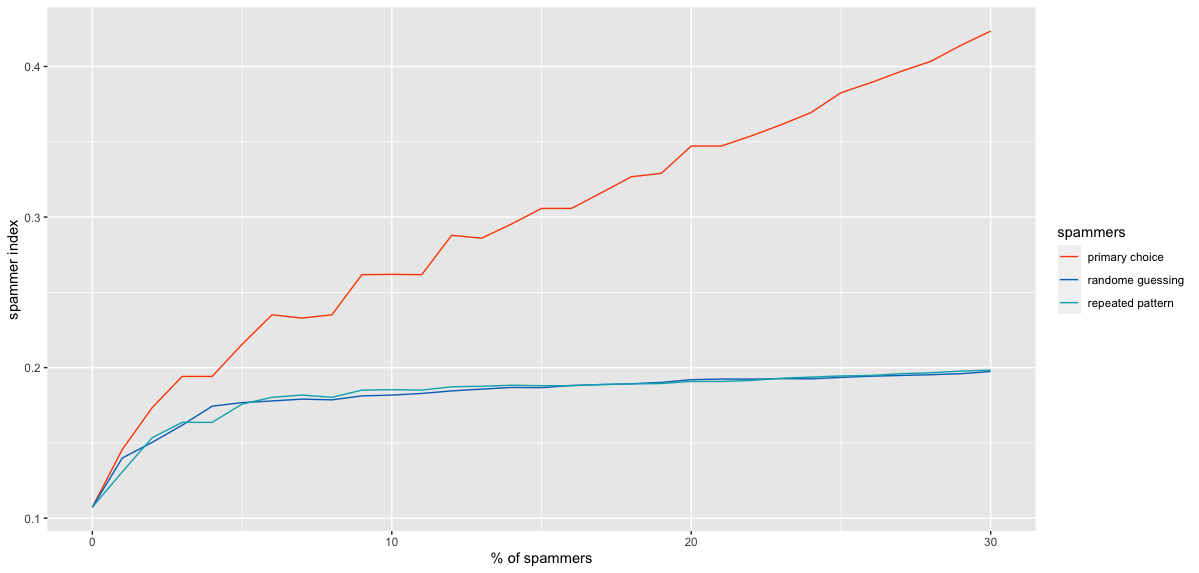}
\end{figure}

To explore the correlation between \emph{Spammer Index} and multiple spamming behaviors, we intentionally added up different percentages of spamming behaviors based on the same amount of credible workers and compared how the \emph{Spammer Index} changed for three different categories. Figure \ref{fig:pct} shows that it is more sensitive to detecting crowd workers whose behaviors belong to \emph{Primary Choice}. Because of this limitation, the \emph{Spammer Index} is used as the first step of data quality evaluation to give us a sense of the \emph{consistency} of the data, which can be considered one indicator of data quality. We will use other statistical methods to detect spamming behaviors based on different response patterns to measure \emph{credibility}. 

\subsubsection{Markov Chains and Metrics based on KL Divergence}
Our methods of identifying spamming behaviors focus on crowd workers' responses. Workers input their responses task by task, which forms sequential discrete data. There are some patterns that distinguish spammers from credible workers within that sequence. Each response is considered a state and it is considered a stochastic process from one state to another state. Markov chain \citep{spedicato2017discrete} is an appropriate method to deal with this discrete space stochastic process. The Markov property is satisfied by this process:  
$$
P(X_{n+1} = s | X_0 = x_0, X_1 = x_1,..., X_n = x_n) = P(X_{n+1} = s|X_n = x_n)
$$
$ X(t) = X_t, t = 0, 1, ..$  and $ S = {0, 1} $  denotes the responses from crowd workers. Because of the memory-less property, a transition probability matrix is estimated to represent the probability of transitioning from one state to another state: $ p_{ij} \geq 0;  p_{ij}(n) = P(X_{n+1} = j |X_n = i) $; for each row in the matrix, they sum up to 1,  $\sum_{j} p_{ij}(n) = 1 $. 

The transition probability matrix estimated from observed data is called the observed transition matrix. The typical spamming matrix is called the target transition matrix. Each type of spammer has a specific target matrix. For example, typical \emph{Random Guessing} spammers have an equal probability of transitioning from one state to another. We define the following three transition matrices: 

\[
\mathbf{P_{pc}} = 
        \begin{blockarray}{ccc}
         & 0   & 1  \\
        \begin{block}{r(rr)}
        0 & 1 & 0  \\
        1 & 1 & 0 \\
        \end{block}
    \end{blockarray}; \qquad
\mathbf{P_{rp}} = 
        \begin{blockarray}{ccc}
         & 0   & 1  \\
        \begin{block}{r(rr)}
        0 & 0 & 1  \\
        1 & 1 & 0 \\
        \end{block}
    \end{blockarray}; \qquad
\mathbf{P_{rg}} = 
        \begin{blockarray}{ccc}
         & 0   & 1  \\
        \begin{block}{r(rr)}
        0 & 0.5 & 0.5  \\
        1 & 0.5 & 0.5 \\
        \end{block}
    \end{blockarray}
\]

$P_{pc}$ represents the target transition matrix of \emph{Primary Choice} spammers (pc), who always have a very high probability of going to the preferred choice no matter what the last choice is. $P_{rp}$ denotes the target transition matrix of \emph{Repeated Pattern} spammers (rp), who have a very high probability of switching their choices. 
$P_{rg}$ is the target transition matrix of \emph{Random Guessing} spammers (rg), whose probability of turning to either choice is 0.5.

Each row of the transition matrix is considered a distribution and a metric is applied to calculate the distance between the i-th row of the observed transition matrix and i-th row of the target transition matrix. Kullback-Leibler divergence (KLD) \citep{belov2011distributions} is commonly used to calculate the similarity between two distributions. Minimizing the Kullback-Leibler divergence is equivalent to maximizing likelihood estimation. Therefore, we propose the average Kullback-Leibler divergence (aKLD) to measure the similarity of observed behavior patterns and typical spamming behaviors.

\begin{equation}
    \textrm{average KLD} = \frac{KLD_{1} + KLD_{2}}{2}
\end{equation}
where $KLD_{i} = D_{KL}(P_{i} ||Q_{i}) = \sum_{x \in X}  P(x) log(\frac{P(x)}{Q(x)})$, i=1, 2 (since, here, we focus on binary cases\footnote{Average KLD can be extended as $\textrm{average KLD} = \frac{\sum (KLD_{k})}{k}$ for ordinal and nominal cases.}); $P_{i}$ is the distribution of i-th row of observed transition matrix and $Q_{i}$ is the distribution of i-th row of the target transition matrix. A small average KLD value means the observed behaviors are similar to the target spamming behaviors. The smaller aKLD a worker has, the more likely the worker is a spammer. We use the simulation method to determine empirical cutoffs for whether a crowd worker's behaviors could be considered spamming. 

Sometimes the estimated transition matrix may exhibit a mixed style that does not belong to any typical defined spamming behavior but their aKLD may be below one of the thresholds that identify spammers. See an example below: 

\[\mathbf{P} = 
        \centering
        \begin{blockarray}{ccc}
         & 0   & 1  \\
        \begin{block}{r(rr)}
        0 & 0.51 & 0.49  \\
        1 & 0.85 & 0.15 \\
        \end{block}
\end{blockarray}; 
\]

In the first row of the transition matrix, the worker may belong to \emph{Random Guessing} behaviors, while the second row of the transition matrix indicates that the worker may be exhibiting \emph{Primary Choice} spamming behavior. However, we do not consider this type of inconsistent spamming behavior. To exclude weak spamming behavior patterns like this, we propose two strategies:  

\begin{itemize}
    \item Apply aKLD to both $KLD_{1}$ and $KLD_{2}$. Only if both $KLD_{1}$ and $KLD_{2}$ are lower than the threshold, we will consider these workers as spammers; 
    \item Use minimum KLD as the threshold\footnote{In the multi-classes context, $\textrm{minimum KLD} = \min{(KLD_{i}, KLD_{2}, ...KLD_{k})}$.} . Minimum KLD is a stricter metric that tends to provide a lower threshold and it finds out the crowd workers whose behaviors present the highest similarity to the characteristics of spamming behaviors. 
\end{itemize}
\begin{equation}
    \textrm{minimum KLD} = \min{(KLD_{1}, KLD_{2})}
\end{equation}

\subsubsection{Deviance Distance}
We introduce a second approach to evaluate the credibility of workers through residual analysis. Residual analysis is a technique to assess regression models' validity and find out the influential data points by model diagnostics. Deviance residuals \citep{smyth2003pearson, pierce1986residuals} are used to evaluate the goodness-of-fit of a generalized linear model (GLM) by comparing the fitted model with a fully saturated model. It is calculated by the sign of actual responses minus predicted responses to quantify the extent to which the probabilities estimated from the fitted model differ from the observed data, enabling the identification of observations that are poorly explained by the model. However, in the case of GLRM, we refer to "the saturated model" as the original model with all observations. 

\begin{equation}
    D = -2(\textrm{log-likelihood}(\phi) - \textrm{log-likelihood}(\phi_{i}))
\end{equation}
where $\phi$ represents parameters of the GLRM with all data, $\phi_{i}$ denotes parameters of the GLRM after deleting observations from worker i. Deviance distance (D) is used to detect the most influential workers whose observations have a significant effect on the overall fit of the model. In other words, the most influential workers are the ones who have different response patterns compared to the rest of the workers. 

Deviance distance equals to the log-likelihood ratio statistic, which follows an asymptotic Chi-squared distribution \citep{wilks1938large}. We pick the significant critical level (0.05) and employ the Chi-squared test to find out the most influential workers who significantly change the model parameters estimation. The degree of freedom $\chi^2$ is equivalent to the difference in dimensionality of $\phi$ and $\phi_{i}$, the number of deleted observations of worker i. Deviance distance is a model-based metric that requires building the generalized linear random effects model correctly to represent the nature of the data, and is identical to the model used to compute the \emph{Spammer Index}. 

In the following sections, we will demonstrate our approaches via a face-verification experiment described in the Introduction using both simulation and real-world scenarios. We further validate the applicability of our approach in multi-categorical scenarios through a simulation study.

\section{Simulation Study}

In this section, we demonstrate our methods for detecting spamming behaviors using simulated data. Using simulation techniques, we can generate synthetic data that mimics various spamming behaviors. This enables us to establish thresholds that aid in identifying and categorizing spamming behaviors accurately. Moreover, the simulation-based approach allows us to validate the robustness and reliability of our detection methods by assessing their performance in a controlled environment.

\subsection{Simulation Procedure}

The data simulation follows the variance decomposition approach, wherein responses from participants consist of multiple random effects, including random effects of tasks, random effects of workers, and random effects of their interactions. Spamming behaviors are simulated by manipulating the random effects, as we regard these effects as intrinsic properties. Since majority voting is the primary strategy to obtain the ground truth via aggregating responses of all participants, we treat the random effects of tasks as the ground truth in simulation. If the random effects of workers and interactions are small enough to make the random effects of tasks dominate, the response is more likely to be correct. Also, random effects of workers represent the workers' individual preferences and random effects of interactions reflect how workers respond to specific tasks. 

Simulation models vary based on different behavioral patterns. Specifically, credible workers have a smaller proportion of variance related to workers and interactions to guarantee that the variance of tasks would contribute the most part of the total variance, which thereby leads to an accuracy range of 0.75 to 0.9 for credible workers. For the \emph{Primary Choice} spamming behavior, given that our tasks involve binary selections, we leverage the results of \citep{schilling1990longest} to calculate the longest run of a binomial experiment. The expected length of the longest one-choice run is $ R_{n} \approx log _{1/p} (n(1-p)) $, where p= 0.5 and n is the number of tasks. When n = 80 for example, we expect the longest run of the same response continuously to be rounded to 6 and the standard deviation of the length of the longest run roughly equals $ (\pi ^2 / 6ln^2 2 + 1/12)^{1/2} =1.873 $. Thus, to approximate a 95\% confidence interval, we simulated the same response to appear in a continuous length of at least 10 times (6+4) as representative of \emph{Primary Choice} spamming behavior.

For the \emph{Repeated Patterns}, we switched the probability of task n and task n+1 by controlling the random effects of interaction. In detail, if they have a 0.8 probability of making one decision for task n, they will have a 0.8 probability of making the opposite decision for task n+1. \emph{Random Guessing} spammers were relatively easier to simulate -- we controlled each decision to have an equal probability of selecting either answer by setting the addition of random effects equal to 0 and the logit of 0 as 0.5.

\subsection{Empirical Average KLD Cutoffs Via Bootstrapping}
Hypothesis testing is employed to determine empirical aKLD cutoffs for detecting spamming behaviors. We control the type I error to 0.05 and simulate 30000 of each type of spammer and 30000 of credible workers. The distributions of aKLD from spammers and credible workers are expected to be different. Our hypothesis is given below: 
$$
H_o: \textrm{the worker behaves credibly}; \ H_a: \textrm{the worker's behavior follows a specific spamming type} 
$$

To test this hypothesis, we use aKLD as the test statistic. When aKLD $<\beta$, we reject the null hypothesis where $ \beta $ is the empirical cutoff that is constructed at the 95 \% confidence level. $\beta$ varies according to different spamming behaviors. In our case, $\beta = \{\beta_{pc}, \beta_{rp}, \beta_{rg}\}$. To be consistent with the data we collected from a real crowdsourcing platform, we experimented with the number of tasks to be 72, 77, or 80 (Table \ref{table:table-1}).

\begin{figure}[ht]
\caption{aKLD distribution in simulation}
\begin{subfigure}{0.32\textwidth}
  \includegraphics[width=\linewidth]{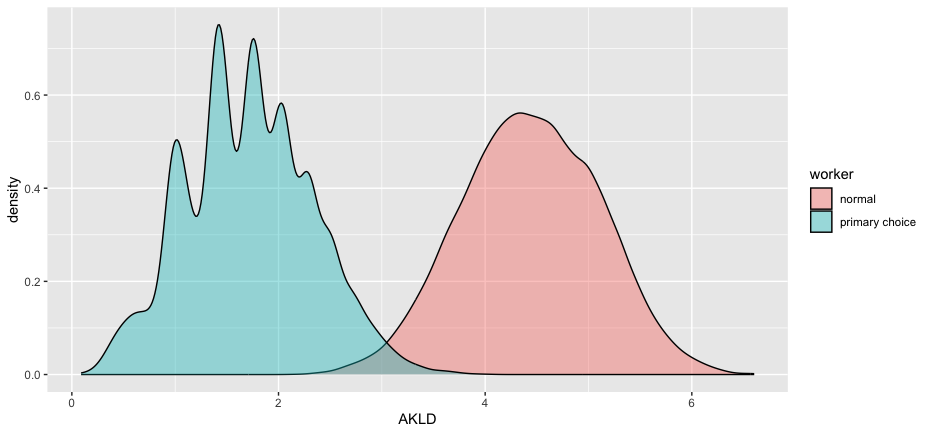}
  \caption{primary choice spammers}
  \label{fig:simulation-pc}
\end{subfigure}
\begin{subfigure}{0.32\textwidth}
  \includegraphics[width=\linewidth]{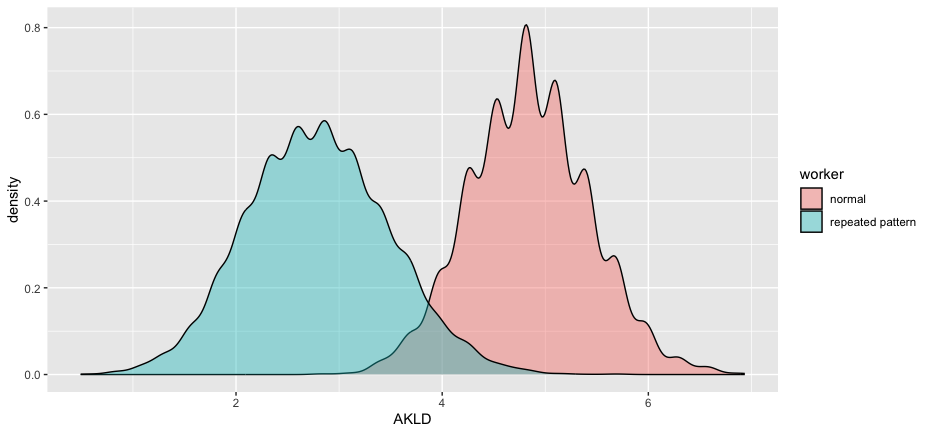}
  \caption{repeated pattern spammers}
  \label{fig:simulation-ps}
\end{subfigure} 
\begin{subfigure}{0.32\textwidth}
  \includegraphics[width=\linewidth]{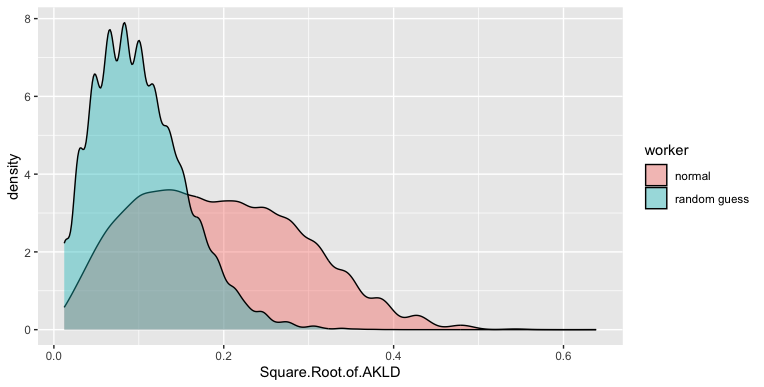}
  \caption{random guessing spammers}
  \label{fig:simulation-rg}
\end{subfigure} 
\label{fig:mKLD-simulation}
\end{figure}

Figure ~\ref{fig:mKLD-simulation} shows aKLD distributions of simulated data, where Figures \ref{fig:simulation-pc} and \ref{fig:simulation-ps} are cases that represent \emph{Primary Choice} spammers and \emph{Repeated Pattern} spammers, respectively. Their distributions depart from that of credible workers. Their type II errors are as small as 0.0044 and 0.0547, respectively. The aKLD cutoffs based on simulation are 3.367109 for \emph{Primary Choice} and 3.924257 for \emph{Repeated Pattern}. 

For \emph{Random Guessing} spammers, since their aKLDs follow an exponential-like distribution, we transform aKLD into its square root form for better comparison. We first use the Kolmogorov-Smirnov test (KS test) \citep{berger2014kolmogorov}, calculating the maximum difference based on the cumulative distribution function (CDF), to compare these two distributions and confirm that they are sampled from two distributions with the p-value of 2.2e-16 and D statistic of 0.4524.

\begin{figure}[!ht]
\caption{Demonstration for effects with imbalanced data and the number of tasks}
\begin{subfigure}{0.4\textwidth}
  \includegraphics[width=\linewidth]{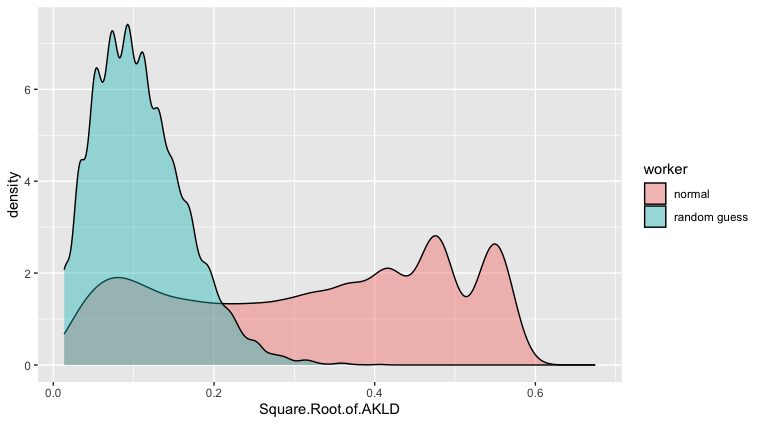}
  \caption{The distributions of aKLD for imbalanced data}
  \label{fig:KL72}
\end{subfigure}
\begin{subfigure}{0.5\textwidth}
  \includegraphics[width=\linewidth]{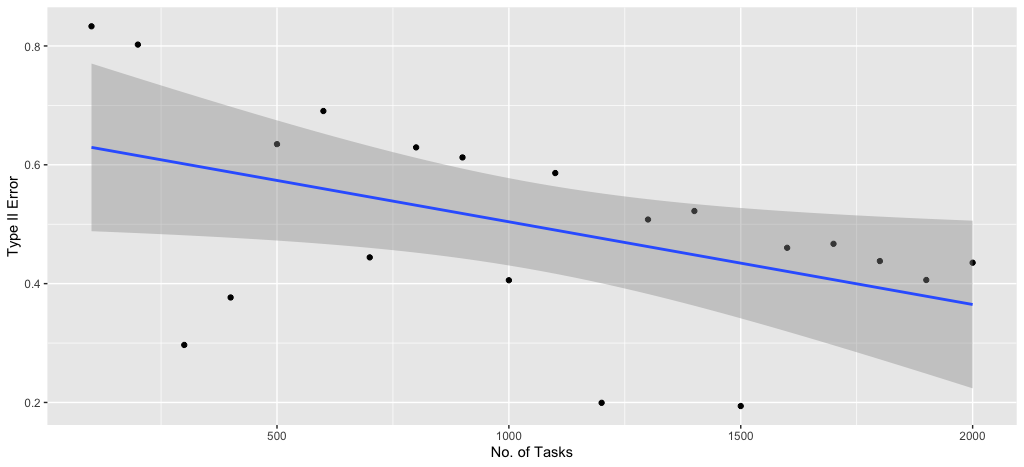}
  \caption{Negative correlation between type II error and NO. of tasks}
  \label{fig:type2}
\end{subfigure} 
\label{fig:klim-2}
\end{figure}

However, the aKLD distribution from \emph{Random Guessing} spammers overlaps with credible workers (see Figure \ref{fig:simulation-rg}). Its type II error is 0.8248 
and aKLD cutoff is 0.002212449. Two reasons for this could be attributed to the remarkable similarity between \emph{Random Guessing} spammers and credible workers, and the small number of tasks. From the response behaviors' perspective, they have a similar pattern. In crowdsourced tasks, tasks are often allocated to workers randomly, resulting in a significant degree of randomness in credible workers' responses. Also, responses from \emph{Random Guessing} spammers are inherent with randomness, which explains their similarity and why the two distributions do not separate well in Figure \ref{fig:simulation-rg}. 
In imbalanced data scenarios (disproportionate distribution for classes of responses), the aKLD demonstrates superior performance (see Figure \ref{fig:KL72}). This improvement is attributed to the reduction of overlap between the distributions, resulting in a decrease in type II error. In such cases, credible workers are less likely to exhibit randomness from tasks with skewed distributions.

On the other hand, a small number of tasks would not accurately reflect the underlying distribution of responses. With the increase in the number of tasks workers deal with, the transition probability matrix from Markov chains can be estimated more precisely. We tested this assumption by increasing the number of tasks up to 2000 for balanced data. We can see the trend is clear in Figure ~\ref{fig:type2}: Type II error decreases as the number of tasks increases, and the type II error observed with 2000 tasks is 0.4352, representing a significant improvement.

From these simulation results, it is evident that the empirical cutoff of aKLD exhibits strong statistical power at the 95\% confidence level for detecting both \emph{Primary Choice} and \emph{Repeated Pattern} spammers. However, in cases of \emph{Random Guessing}, while we can confidently identify workers flagged by the cutoffs as at least 95\% likely to be spammers, we may overlook many other potential \emph{Random Guessing} spammers.

In the subsequent analysis, we incorporate time as an auxiliary factor to filter the workers determined by aKLD, as time plays a decisive role in classifying spammers \citep{wang2020collueagle}. Assuming that spammers are motivated to complete tasks as quickly as possible with little to no human attention on the tasks being asked of them, we employ a filtering approach to identify crowd workers whose average task completion time falls below one standard deviation below the mean. Additionally, spending less time on particularly difficult tasks indicates carelessness or unreliability of workers' responses, potentially leading to lower accuracy in results, even if spamming behavior is not intentional.

\subsection{Deletion Analysis}
To assess the deletion analysis approach, we simulated 10\% spammers out of the total workers with the same simulation strategy, which contains 108 credible workers and 12 spammers, including 4 \emph{Primary Choices}, 4 \emph{Repeated Patterns}, and 4 \emph{Random Guessers}. The IDs of spamming workers are from 1 to 12. Each worker is assigned to work on 80 different tasks. The dimension of simulated data is 120*80 = 9600. The data quality (\emph{Spammer Index}) of credible workers is  0.01272426 and after involving spammers, this value increases to 0.0607043.

\begin{figure}[ht]
\caption{Deviance Distance}
\begin{subfigure}{0.45\textwidth}
  \includegraphics[width=\linewidth]{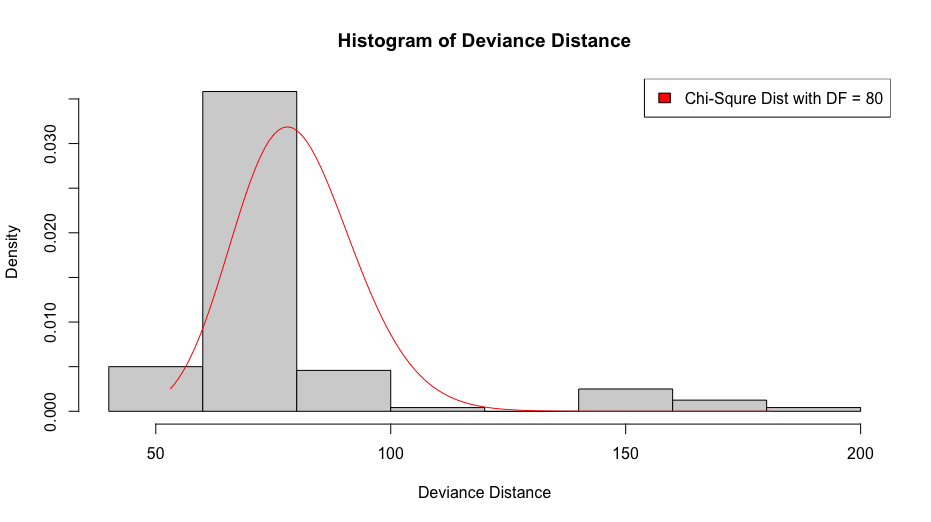}
  \caption{Deviance distance follows Chi-squared distribution}
  \label{fig:deviance}
\end{subfigure}
\begin{subfigure}{0.5\textwidth}
  \includegraphics[width=\linewidth]{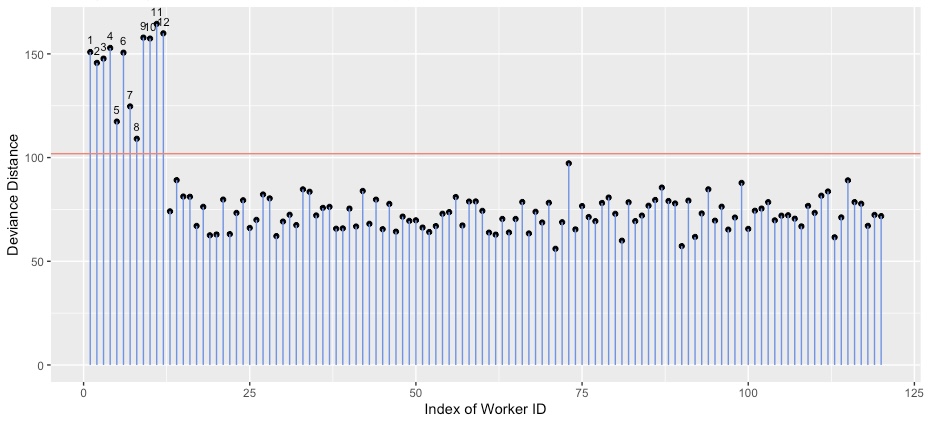}
  \caption{Deviance distances for worker i removed, cutoff (p-value) based on a Chi-squared test with a significance level of 0.05 and df = 80 }
  \label{fig:deltion-simulation}
\end{subfigure} 
\label{fig:dd}
\end{figure}

We generate the histogram of deviance distance for the simulation data and fit a Chi-squared distribution with 80 degrees of freedom (the red curve in Figure ~\ref{fig:deviance} ) to verify the Chi-squared distribution of deviance distance. Figure \ref{fig:deltion-simulation} is the result after performing deletion analysis. The red line is the distribution of the Chi-squared test with 80 degrees of freedom for a significance level of 0.05. The deviance distances of the first 12 workers all exceeded the cutoff, indicating the successful detection of all spammers.

One drawback of deletion analysis is its heavy computational burden when dealing with large datasets, as it requires iterating over all observations from each worker. Additionally, we will obtain little information about the type of spamming behavior. However, by comparing the value of deviance distance, the deletion method appears relatively insensitive to \emph{Primary Choice} spamming. This observation is evident in Figure \ref{fig:deltion-simulation}, where the deviance distances associated with \emph{Primary Choice} spammers (Worker No. 5 to No. 8) are smaller compared to those of the other two spammer types. It is recommended that applying both methods to measure \textit{credibility} would be beneficial.

\begin{figure}[ht]
\caption{Deviance distance in simulated multi-class response data}
\begin{subfigure}{0.45\textwidth}
  \includegraphics[width=\linewidth]{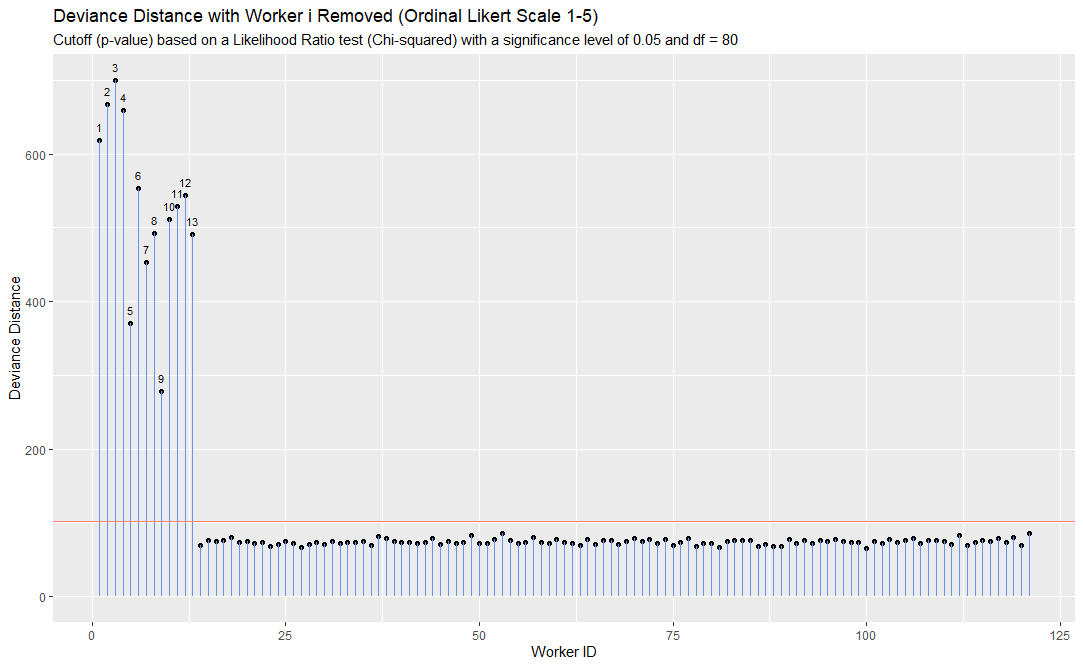}
  \caption{deletion analysis for ordinal data}
  \label{fig:ordinal}
\end{subfigure}
\begin{subfigure}{0.44\textwidth}
  \includegraphics[width=\linewidth]{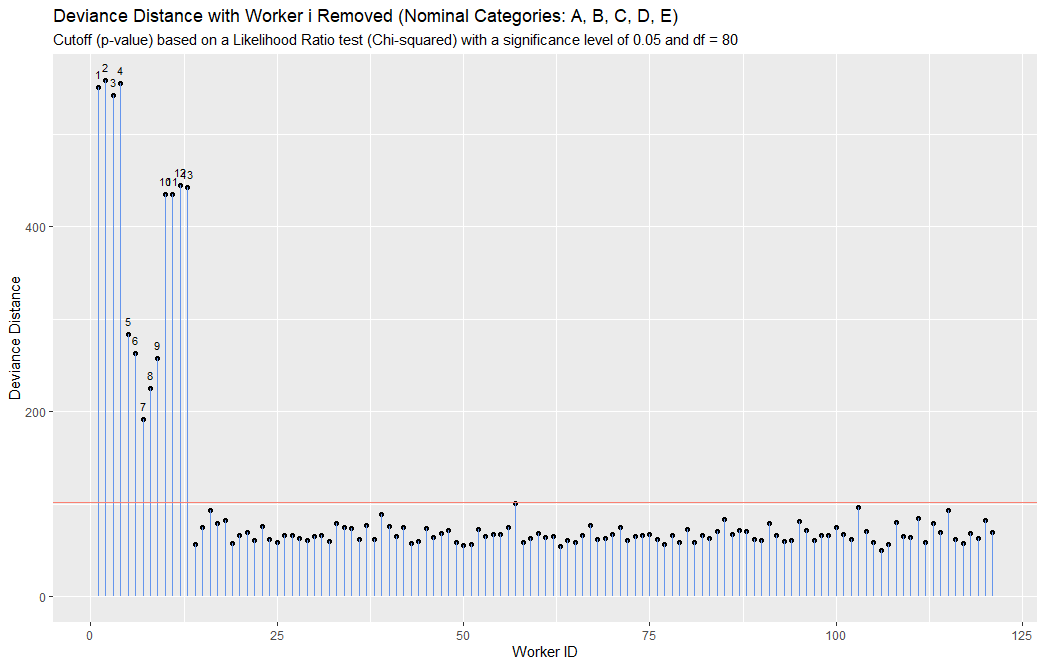}
  \caption{deletion analysis for nominal data}
  \label{fig:nominal}
\end{subfigure} 
\label{fig:multi-class}
\end{figure}

To validate our approach in a multi-class context, we simulate ordinal and nominal response data and perform deletion analyses \footnote{We use the ``ordinal'' package \citep{ordinal} for the ordinal cases and ``mclogit'' \citep{elff2022mclogit} for the nominal cases. Other packages, such as ``brms'' \citep{burkner2017brms}, can handle both scenarios.}. The simulation strategies are outlined in Appendix B. Figure \ref{fig:multi-class} illustrates that the deviance distances for simulated spammers (worker IDs 1-12) are significantly larger than those of normal workers. Furthermore, using the p-values derived from a likelihood-ratio (Chi-squared) test as the threshold effectively identifies all spammers.

\section{Experiment}

In this section, we present an experiment mentioned in the introduction to validate the effectiveness of our methods in measuring crowd workers' \emph{consistency} and \emph{credibility}. We illustrate the application of our proposed metrics to assess data quality in a real-world scenario and filter participants using auxiliary information, the time spent on tasks, to identify spammers. We use the accuracies of the workers' responses in our experimental tasks (in which the ground truth is known by the experimenters) to serve as the standard in our validation.

\begin{wrapfigure}{R}{0.33\textwidth}
\caption{\label{fig:fw} the face-verification interface called Facewise}
\centering
\includegraphics[width=0.35\textwidth]{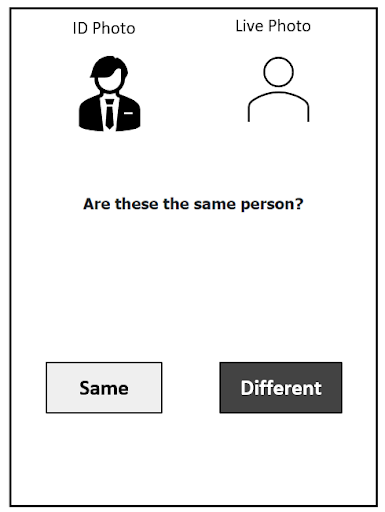}
\small\textit{Note: Participants can see the real image pairs in the experiment; we use an icon here for privacy purposes.}
\end{wrapfigure}

We designed a set of face-verification tasks to gather responses from crowd workers. In these tasks, workers are instructed to imagine themselves as Transportation Security Officers (TSOs) responsible for the ID verification process at airport checkpoints. Their task is to compare two similar face images and determine whether the image pair depicts the same person(Figure \ref{fig:fw}). Their responses are logged as either "Same" or "Different". 
\subsection{Data Collection in Crowdsourcing and from Experts}
We selected two representative crowdsourcing platforms: Amazon Mechanical Turk (MTurk) and Prolific. MTurk \citep{mturk} is one of the earliest and largest crowdsourcing platforms with a vast pool of workers from diverse backgrounds. While Prolific\citep{prolific}, a newer platform used primarily for academic research, has experienced rapid growth in the past few years. Both platforms offer monetary compensation. Previous research \citep{peer2022data} states that the data quality collected from Prolific exceeds that in MTurk. Therefore, we expect data collection via Prolific to be of better quality and involve fewer spamming behaviors. We recruited 160 participants from MTurk, each tasked with completing 77 tasks, and 138 participants from Prolific, each assigned 72 tasks.

Furthermore, to facilitate a comparison with the data quality in crowdsourcing, we enlisted 152 subject matter experts, U.S. Transportation Security Officers (TSOs) from three major U.S. airports in Phoenix, Las Vegas, and San Diego. Each of TSOs agreed to participate in our experiment without financial rewards to complete 80 face-matching tasks. We assume that data collection from TSOs exhibits high quality because of their expertise in face-matching tasks. However, this does not necessarily imply that they are consistently motivated and focused throughout tasks to provide quality data without any spamming in the light that they could feel bored due to the difference between our experiment and the real tasks they face daily and they are volunteers in the absence of monetary incentives. 
Table ~\ref{table:table-1} provides details of the data and data quality in these three experiments. 

\begin{table}[h!]
\centering
\caption{Datasets Description}
\begin{tabular}{ c c c c}
 \hline
 Platform  & Number of Workers & Number of Tasks & Spammer Index\\\hline
 MTurk     & 160    &  77  &   0.166\\ 
 Prolific  & 138    &  72  &   0.065 \\ 
 Airports  & 152    &  80  &   0.079\\ \hline
\end{tabular}
\label{table:table-1}
\end{table}

The generalized linear random-effects model is built by R programming language and the package "lme4" \citep{lme4}. The model structure in R code is below:

\begin{lstlisting}[language=R, basicstyle=\scriptsize\ttfamily, linewidth=1.2\textwidth]
Model = glmer(response ~ (1|workerid ) + (1|taskid) + (1|workerid:taskid),     
        data = data, family = binomial, control = glmerControl(
        optimizer = "bobyqa", optCtrl = list(maxfun = 1e6)), nAGQ = 1)        
\end{lstlisting}

\subsection{aKLD Performance}
Applying the aKLD method requires empirical thresholds that are derived from simulated data based on the descriptions in Table \ref{table:table-1}. We implement the first strategy discussed in the Proposed Metric section: Apply aKLD to both $KLD_{1}$ and $KLD_{2}$. Only those individuals with both $KLD_{1}$ and $KLD_{2}$ values smaller than the threshold would be deemed to pass our selection criteria. Table \ref{table:table-3} shows the results. Workers' accuracy is calculated through all tasks as the standard to validate the performance of the Markov chain \& aKLD method. 
Two scales for accuracy evaluation are employed here: lower than the mean, which is lower than 50\% of participants, and lower than one standard deviation (SD) below the mean, which is lower than 84\% of participants. 

\begin{table}[h]
\caption{Spammers detection based on aKLD}
\centering
\renewcommand{\arraystretch}{1.5}
\resizebox{\textwidth}{!}{%
\begin{tabular}{@{}*{8}{c}@{}}
\toprule
  Platform 
  & \thead{Number of \\ total spammers}
  & \thead{Number of \\ spammers for \\ Primary \\ Choice}
  & \thead{Number of \\ spammers for \\ Repeated \\ Pattern}
  & \thead{Number of \\ spammers for \\ Random \\ Guessing}
  & \thead{Number of \\ spammers whose \\ accuracy lower \\ than the mean}
  & \thead{Number of \\ spammers whose \\ accuracy lower \\ than 1 SD \\ below the mean}
  \\
\midrule
 MTurk    & 29 (18)  &  18 (12) &  6 (5) & 5 (1)  & 22 (15) & 11 (7)\\  
 Prolific & 10 (5)   &  5 (2) &   3 (3)  & 2 (0) & 5(2) & 3(1) \\
 Airports & 13 (8)   &  8(3)  &   3(3)  &  2(2)  & 4(3) & 4(3) \\ 
\bottomrule
 \end{tabular}%
}
\newline
\newline
\begin{tablenotes} 
\textit{Note:}
The number within parentheses denotes the amount of identified spammers after applying time filtering: whose average time on tasks is below the mean. 
\end{tablenotes}
\label{table:table-3}
\end{table}

Given that spammers are assumed to spend less time accomplishing tasks, we can identify those whose average task completion time is below the overall average to reduce the false positive error. The values within parentheses in Table \ref{table:table-3} represent the numbers of spammers detected for each dataset after applying the filter based on task completion time \textemdash specifically, those whose average time on tasks is below the overall average.

Table \ref{table:table-3} shows that the number of detected spamming behaviors is highly correlated with the \emph{Spammer Index} in Table \ref{table:table-1}, where we found some interesting trends. 1) When the \emph{Spammer Index} is higher, we tend to observe a larger count of spammers. Simplistic spamming behaviors (\emph{Primary Choice} and \emph{Repeated Patterns}) are relatively common.  
2) As the \emph{Spammer Index} increases, the correlation between detected spammers and lower accuracy becomes more apparent. The data collected from MTurk has the worst data quality of these three sources, and among the spammers detected, 74.86\% (83.33\% after time filtering) of their accuracies are lower than the average, and 37.93\% (38.89\% after time filtering) are lower than one standard deviation below the average. 
In contrast, the data from Prolific exhibits the highest data quality among these three datasets, with the fewest spammers detected. Among them, 50\% (40\% after time filtering) of their accuracies fall below the average, and 30\% (20\% after time filtering) is lower than one standard deviation from the average. For the data collected from TSOs,  30.77\% (37.5\% after time filtering) of the accuracies of spammers fall below the average, while 30.77\% (37.5\% after time filtering) are lower than one standard deviation from the average, respectively.

\subsection{Deletion Analysis Results}

\begin{figure}[!h]
\caption{Deletion Analysis from three sources}
\begin{subfigure}{0.32\textwidth}
  \includegraphics[width=\linewidth]{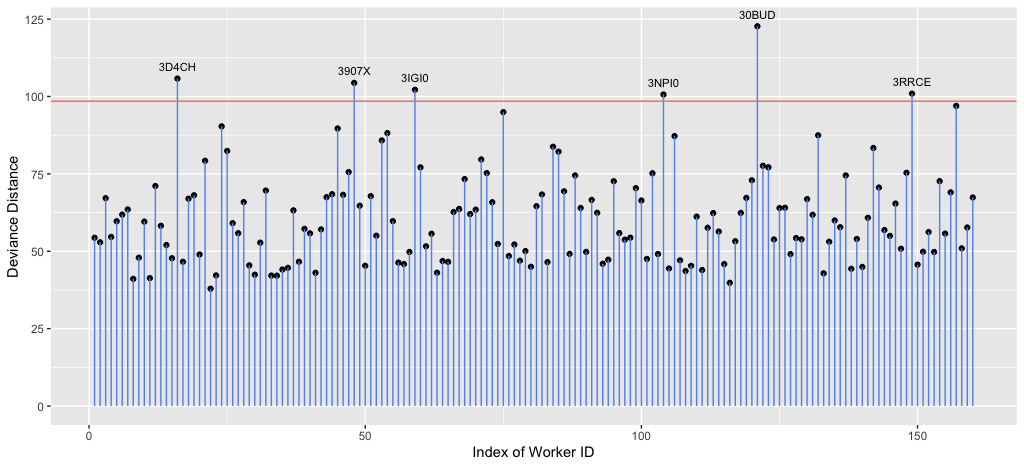}
  \caption{Deviance distances in MTurk, cutoff based
on a Chi-squared test with df = 77}
  \label{fig:deletion-mt}
\end{subfigure}
\begin{subfigure}{0.32\textwidth}
  \includegraphics[width=\linewidth]{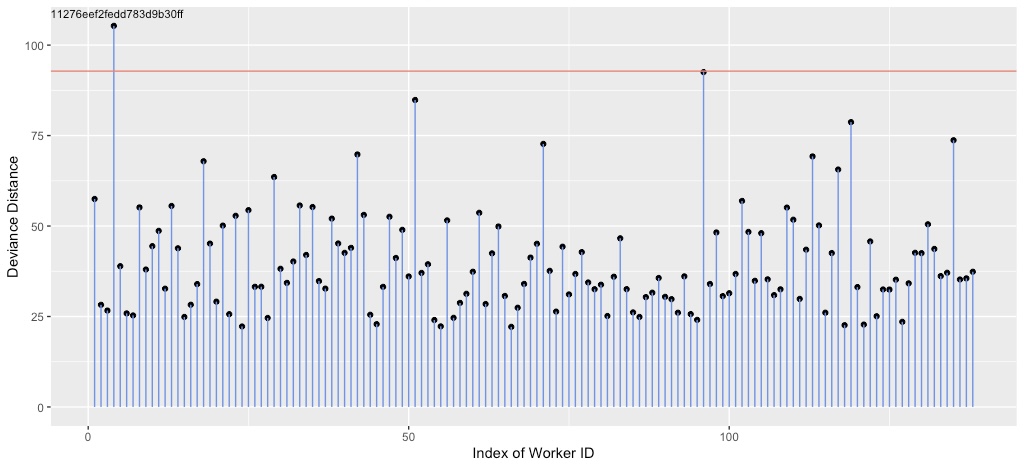}
  \caption{Deviance distances in Prolific, cutoff based
on a Chi-squared test with df = 72}
  \label{fig:deletion-pl}
\end{subfigure} 
\begin{subfigure}{0.32\textwidth}
  \includegraphics[width=\linewidth]{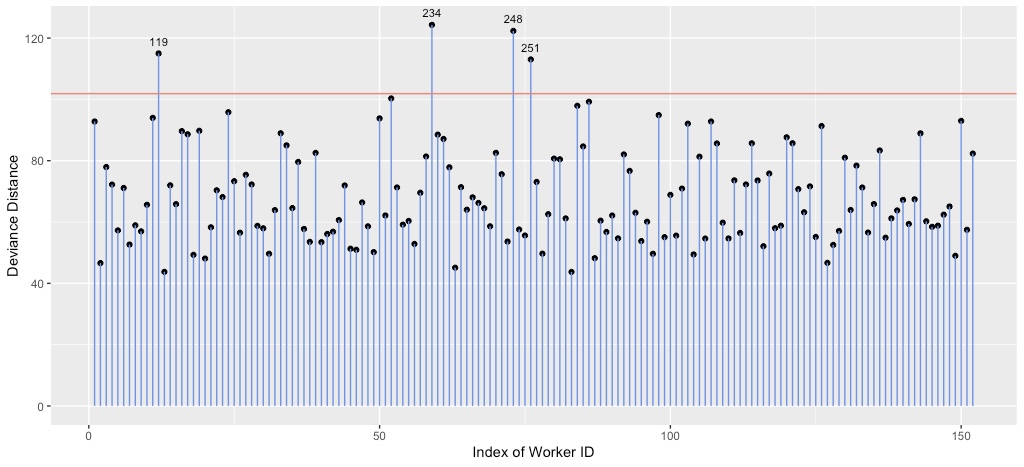}
  \caption{Deviance distances from airports, cutoff based
on a Chi-squared test with df = 80}
  \label{fig:deletion-fw}
\end{subfigure} 
\label{fig:deletion-3p}
\end{figure}

To implement deletion analysis, we employ the generalized linear random effects model, sharing the same underlying structure as the model used to calculate the \emph{Spammer Index}, to compute the deviance distance. Deletion analysis is carried out based on the worker ID. In Figure \ref{fig:deletion-3p}, the worker IDs above the red line represent potential spammers who exceed the thresholds set by the Chi-squared test (0.05 significance level). In the MTurk data, we identify 6 crowd workers whose response patterns are significantly different from the rest. 1 potential spammer is detected in the Prolific data, while 4 spammers are found in the airport data. The result is consistent with the assumption that a larger \emph{Spammer Index} indicates more potential spamming behaviors in the data.    

Six potential spammers are detected in MTurk(Figure \ref{fig:deletion-mt}), the accuracies of whom all fall below the average and five out of six are lower than one standard deviation from the average (Table \ref{table:table-2}). Only one potential spammer is observed above the cutoff in Prolific (Figure \ref{fig:deletion-pl}). The number of spammers in the airport datasets ranks between MTurk and Prolific (Figure \ref{fig:deletion-fw}), all of whose accuracies are lower than 1 standard deviation below the average (Table \ref{table:table-2}). Compared to the aKLD method, deletion analysis tends to uncover fewer instances of spamming behaviors, but those detected are often more severe. Individuals identified by model-based deviance distance may not easily fit into specific categories of spammers, but a significant majority exhibit notably low accuracy. 

\begin{table}[!h]
\centering
\caption{Deletion Analysis}
\begin{tabular}{@{}*{15}{c}@{}}
\toprule
  Platform 
  & \thead{Number of total spammers}
  & \thead{Number of spammers \\whose accuracy lower \\than the mean}
  & \thead{Number of spammers \\ whose accuracy lower \\ than 1 SD \\ below the mean}\\
\midrule
 MTurk    & 6    &  6  &   5\\  
 Prolific & 1    &  1  &   1\\ 
 Airports & 4    &  4  &   4\\ 
\bottomrule
 \end{tabular}
\newline
\newline
\label{table:table-2}
\end{table}

\subsection{Categorize Potential Spammers}

In the two approaches described above, we categorize and evaluate potential spammers into three levels: "High risk", "Moderate risk", and "Undetermined risk" (Table \ref{table:category}). These classifications are based on three perspectives of spamming behaviors: behavior patterns, average time on tasks, and accuracy. Scores in each dimension can take on values of 0, 0.5, or 1. Specifically, if a worker's behavior aligns with one of the specific typical spamming behaviors identified through examination of the transition probability matrix, they will receive 0.5 and 0 otherwise. Workers will receive an additional score of 0.5 if their response patterns match typical spamming behaviors. For the average time and accuracy, workers will get a score of 0 if their average time or accuracy is above the mean, 0.5 if it is below the mean, and 1 if it is lower than one standard deviation below the mean. Finally, the scores from these three dimensions are summed, and workers are categorized as follows:

\begin{itemize}
    \item "High risk": if the total score is greater than or equal to 2.5. Workers or participants whose scores fall on or above this threshold are most likely spammers and are recommended for exclusion.
    \item "Moderate risk": if the total score falls between 1.5 and 2. Workers or participants whose scores fall within this range are more likely to be spammers and may be worth manual inspection and judgment. 
    \item "Undetermined risk" if the total score is between 0 and 1. Workers or participants with these scores present ambiguity in classification because of equal odds of being true credible or spammers. 
\end{itemize}

\begin{table}[h]
\caption{Categorizing Potential Spammers and Comparing Spammer Index with Data Quality Metrics}
\renewcommand{\arraystretch}{1.5}
\begin{flushright} 
\resizebox{\textwidth}{!}{
\begin{tabular}{ c c c c c | c c c }
 \hline
  Platform & Spammer Index & High risk & Moderate risk & Undetermined risk & Fleiss' Kappa & ICC (C,1) & ICC (fixed error) \\
 \hline
 MTurk    &  16.6\%  & 9   &  17  &   9  & 0.435 & 0.438 & 0.111 \\ 
 Prolific &  6.5\%   & 1    &  3   &   6  & 0.638 & 0.641 & 0.052 \\ 
 Airports &  7.9\%   & 3    &  4   &   9  & 0.426 & 0.429 & 0.048 \\ 
 \hline
\end{tabular}
}
\end{flushright}
\label{table:category}
\end{table}

The right three columns of Table \ref{table:category} present the commonly-used reliability measures, and they are compared with our proposed \textit{Spammer Index}. Fleiss' Kappa and ICC(A,1) \citep{liljequist2019intraclass} are computed using the "irr" package in R \citep{gamer2012package}. Fleiss' Kappa extends Cohen's Kappa to accommodate multiple raters, while ICC(A,1) measures absolute agreement in a two-way random-effects model, considering both rater and subject variability in continuous ratings. Both measures are highly sensitive to missing values. Without removing NA values for Prolific dataset, Fleiss' Kappa is initially 0.00374 and ICC(A, 1) is 0.0145. ICC (fixed error) is computed by adding a fixed term, the variance of the standard logistic distribution (3.29), to the denominator of Equation \ref{eq2}, as discussed in Section 3 \citep{grilli2007multilevel}. For Fleiss' Kappa and ICC(A,1), higher values indicate better agreement, whereas, for the \textit{Spammer Index} and ICC (fixed error), lower values are preferable. First, neither Fleiss' Kappa nor ICC(A,1) can differentiate the data quality between the MTurk and Expert datasets, even assigning higher reliability scores to MTurk data. Additionally, ICC (fixed error) is problematic, as its fixed variance term remains constant regardless of the total variance, leading to an inversion of values between the Airports dataset and Prolific. This result is unexpected, given that Prolific has noticeably fewer detected spammers. Through comparison, the \textit{Spammer Index} proves advantageous by proportionally reflecting the extent of spamming behaviors present in the collected dataset.

\subsection{Crowd Workers' Demographics}
We acknowledge that demographic information plays an important role in online data collection, and numerous studies have examined its impact. For example, \cite{huff2015these} found that MTurk underrepresents older African Americans compared to national surveys like CCES while effectively recruiting young Hispanic and Asian respondents. \cite{difallah2018demographics} provides insights into various demographic aspects of MTurk workers, including age distribution, gender, marital status, household size, and income levels.

\begin{figure}[ht]
\caption{Demographic Information in Amazon MTurk}
\begin{subfigure}{0.45\textwidth}
  \includegraphics[width=\linewidth]{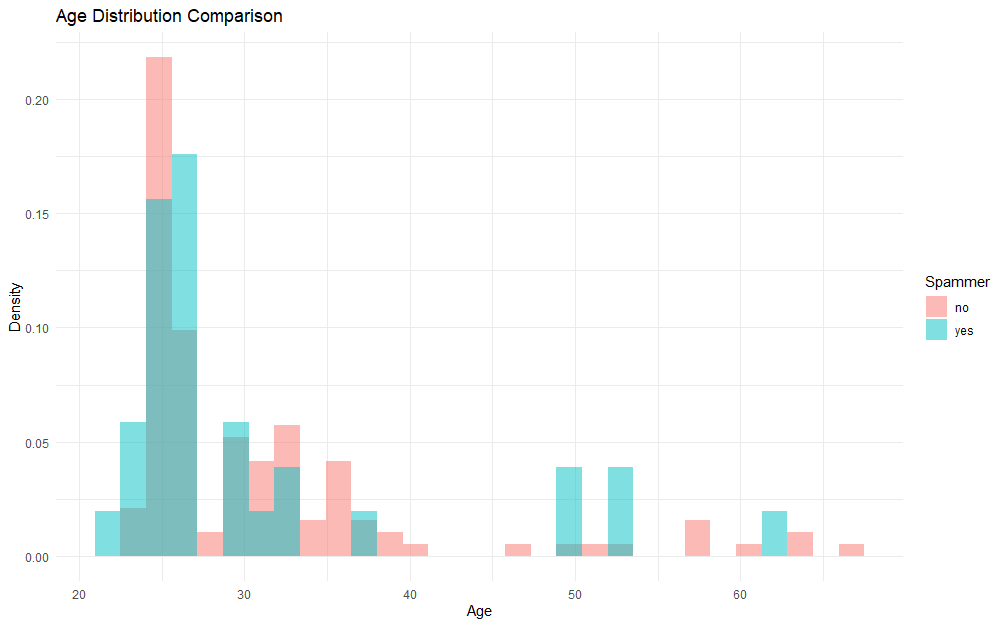}
  \caption{Age distribution in MTurk}
  \label{fig:mt_age}
\end{subfigure}
\begin{subfigure}{0.44\textwidth}
  \includegraphics[width=\linewidth]{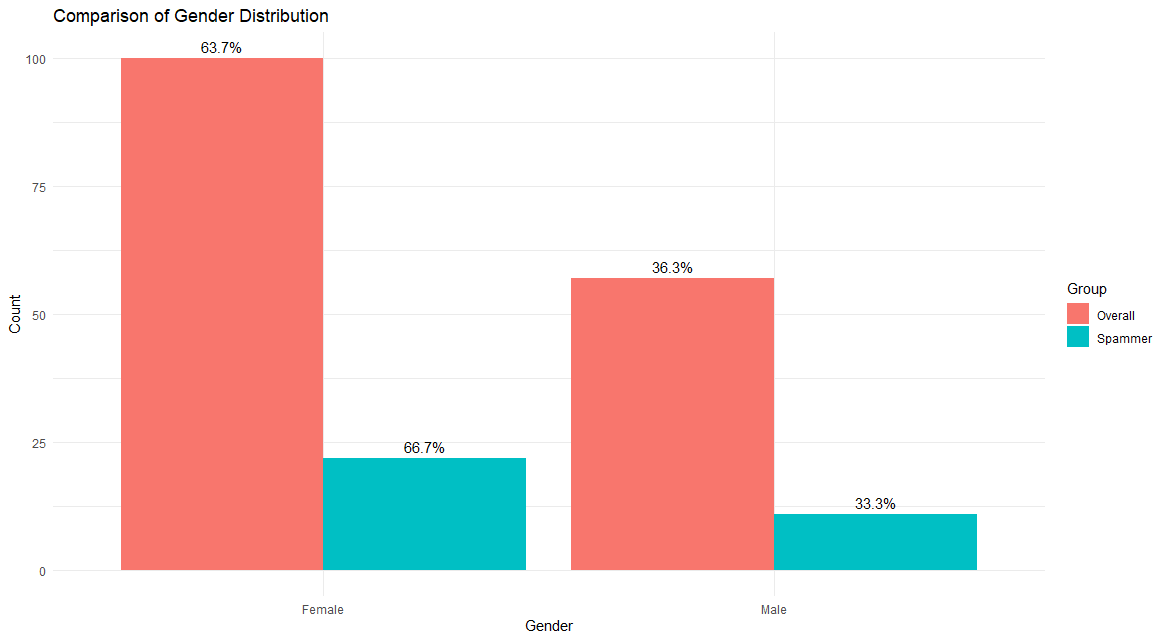}
  \caption{Gender distribution in MTurk}
  \label{fig:mt_gender}
\end{subfigure}

 \vspace{50pt}

\begin{subfigure}{0.45\textwidth}
  \includegraphics[width=\linewidth]{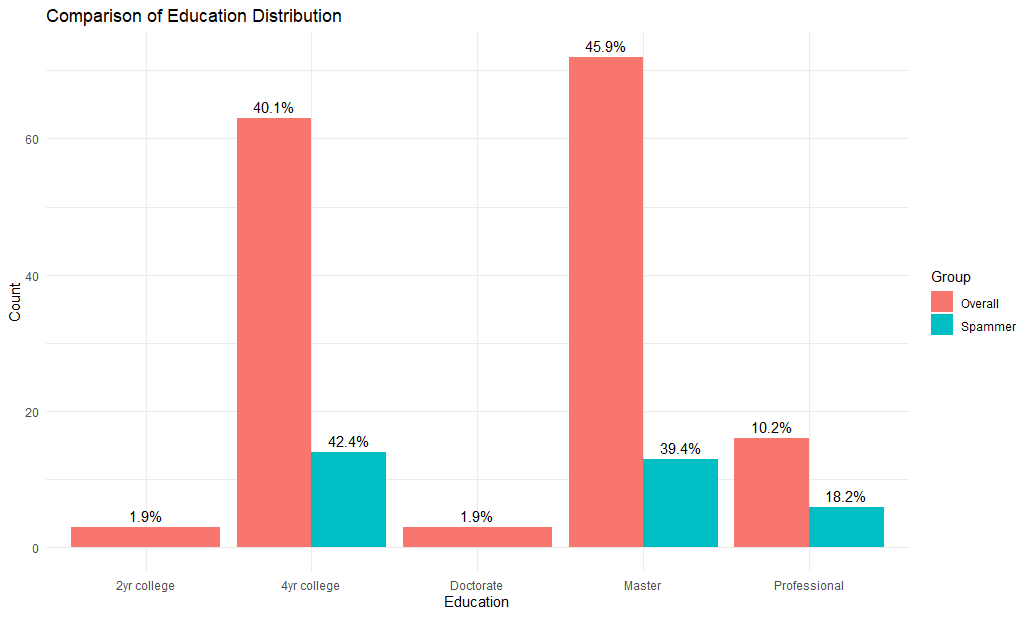}
  \caption{Education distribution in MTurk}
  \label{fig:mt_education}
\end{subfigure}
\begin{subfigure}{0.44\textwidth}
  \includegraphics[width=\linewidth]{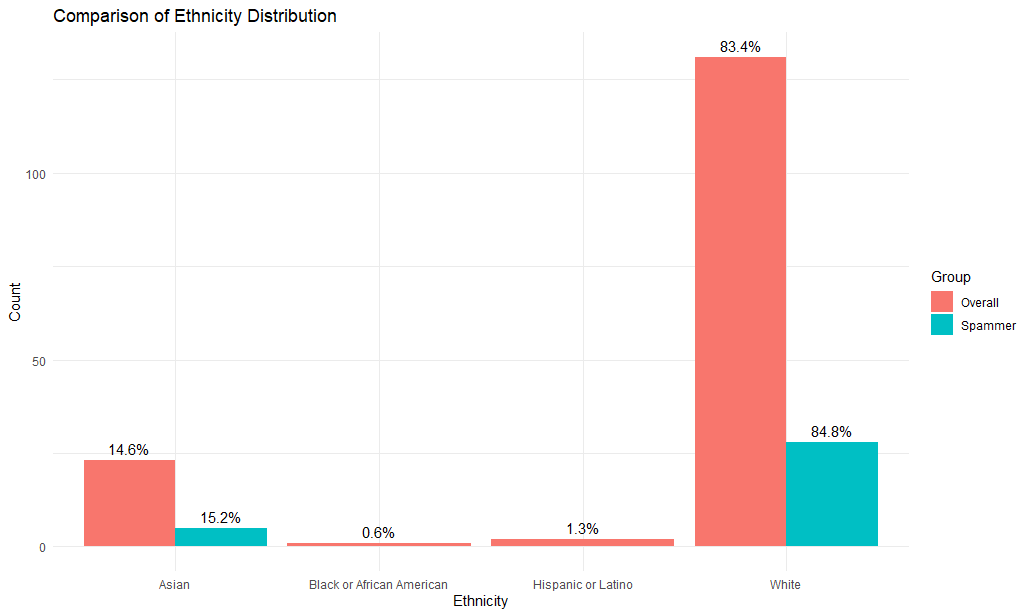}
  \caption{Ethnicity distribution in MTurk}
  \label{fig:mt_ethnic}
\end{subfigure} 
\label{fig:demo-mt2}
\end{figure}

\begin{figure}[ht]
\caption{Demographic Information in Prolific.co}
\begin{subfigure}{0.45\textwidth}
  \includegraphics[width=\linewidth]{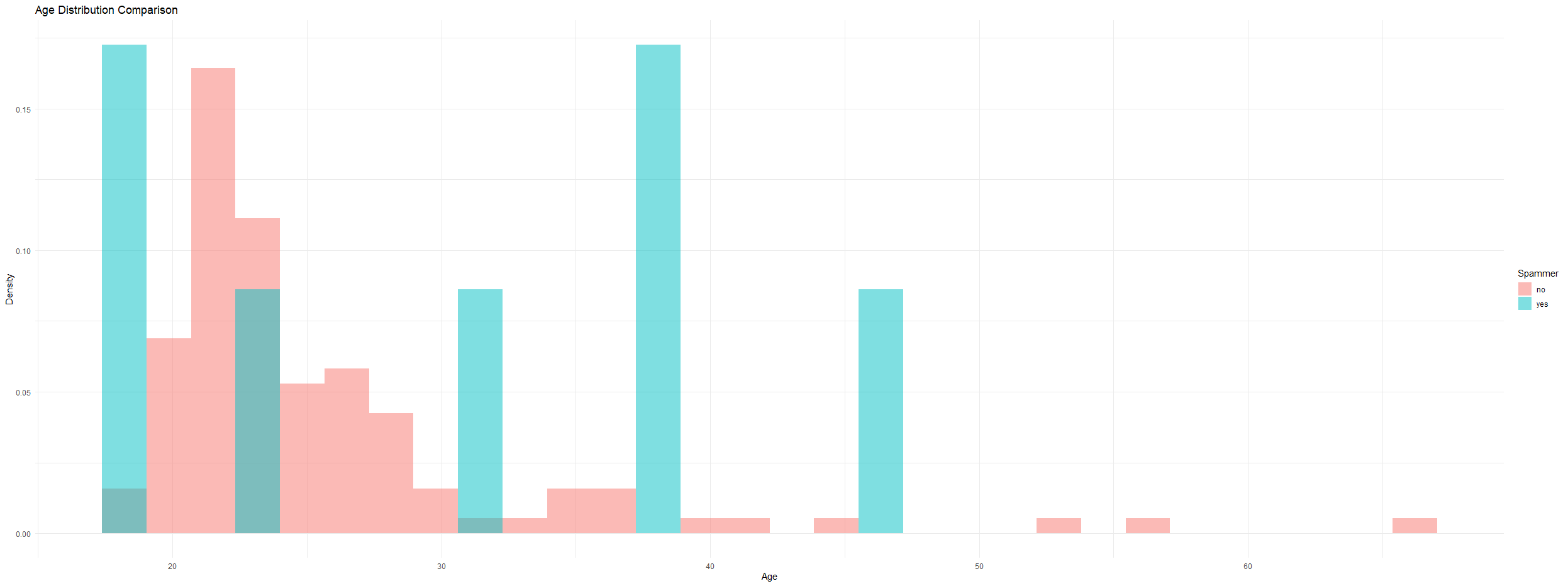}
  \caption{Age distribution in Prolific}
  \label{fig:pl_age}
\end{subfigure}
\begin{subfigure}{0.44\textwidth}
  \includegraphics[width=\linewidth]{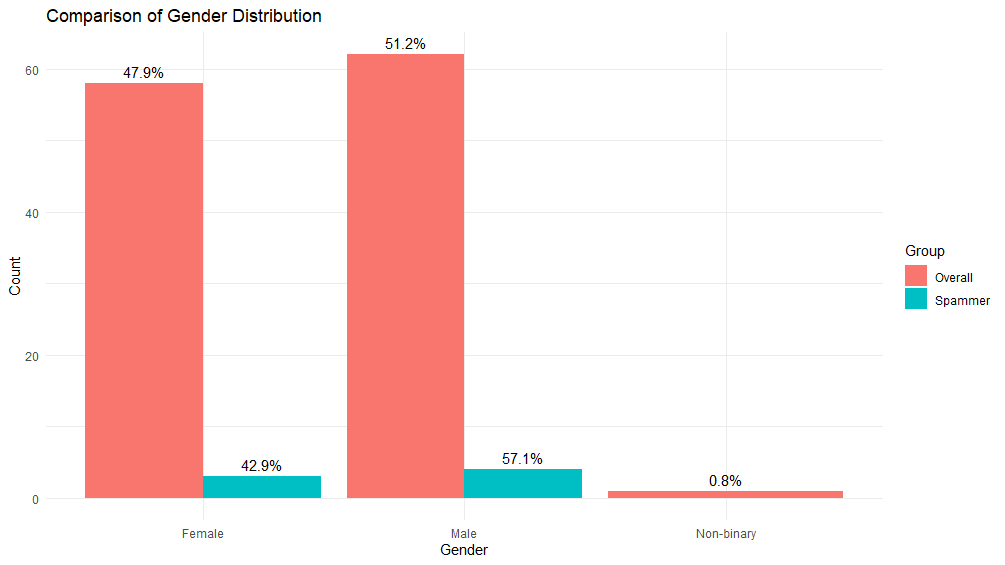}
  \caption{Gender distribution in Prolific}
  \label{fig:pl_gender}
\end{subfigure}

 \vspace{50pt}

\begin{subfigure}{0.45\textwidth}
  \includegraphics[width=\linewidth]{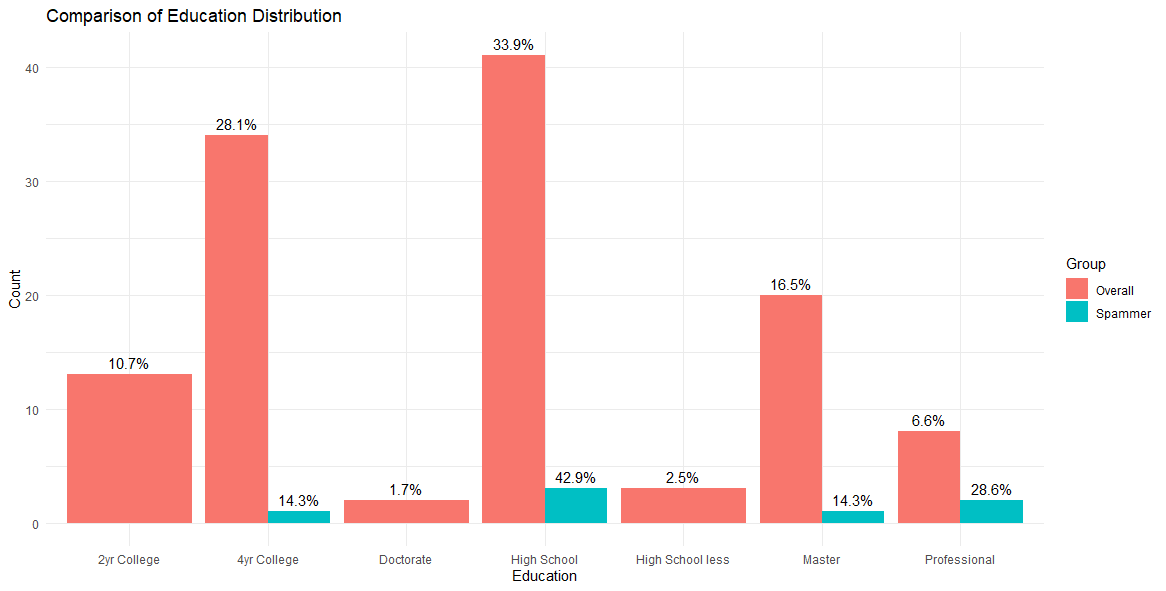}
  \caption{Education distribution in Prolific}
  \label{fig:pl_education}
\end{subfigure}
\begin{subfigure}{0.44\textwidth}
  \includegraphics[width=\linewidth]{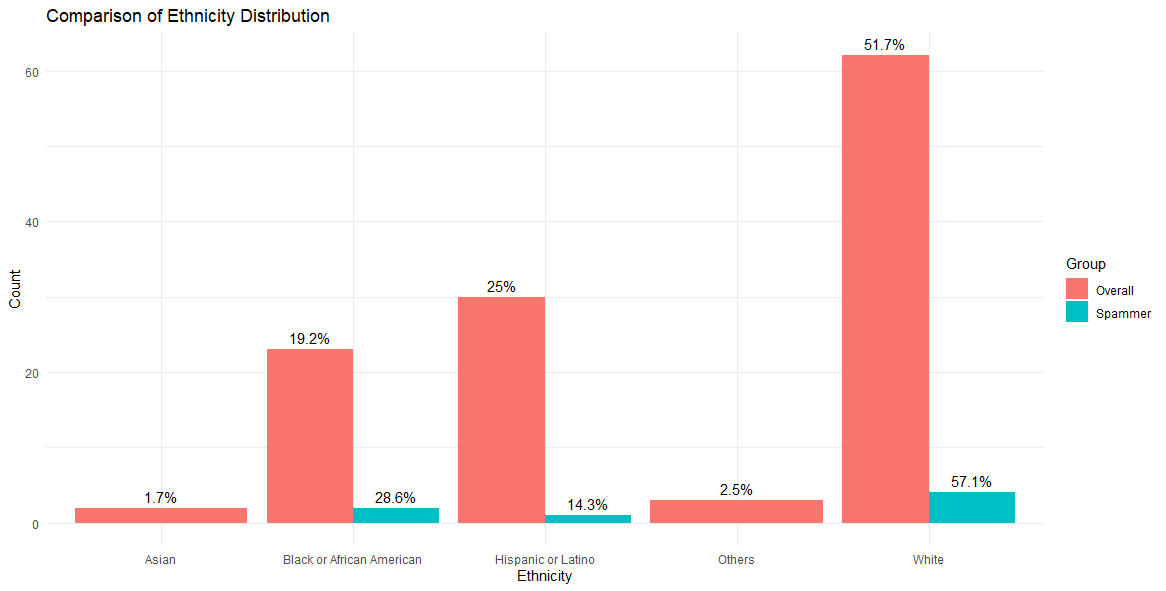}
  \caption{Ethnicity distribution in Prolific}
  \label{fig:pl_ethnic}
\end{subfigure} 
\label{fig:demo-pl2}
\end{figure}

The results in our experiments indicate that older participants (age >40 on Prolific and age >50 on MTurk) are disproportionately more likely to appear in the spammer group compared to younger workers (Figure \ref{fig:mt_age} and Figure \ref{fig:pl_age}). Additionally, possessing an advanced degree does not necessarily correlate with higher reliability (Figure \ref{fig:mt_education} and Figure \ref{fig:pl_education}). Moreover, we do not observe a significant difference between the overall distribution and the spammer distribution concerning gender and ethnicity factors on both Amazon MTurk and Prolific.co (See figures in Appendix D).

\section{Discussion}

\subsection{Empirical Guidelines}

We propose an empirical set of guidelines and recommendations (see Figure \ref{fig:flowchart}) for measuring \emph{consistency} and \emph{credibility}, and consequently, identifying non-credible behaviors in online experiments. First, it is recommended that researchers use a variance decomposition method, such as the GLRM, for initial model fitting and for evaluating the \emph{consistency} of workers on the same set of tasks. This step produces a \emph{Spammer Index}, or SI. Based on simulations and case studies, we found a threshold of $SI\geq0.10$ to be a reasonable criterion for estimating that there are potentially $SI\times N$ workers (where $N$ is the total number of workers) that may exhibit spamming behavior. The use of GLRM at the early stage of data analysis serves two purposes -- 1) it can determine the presence of potential spammers in the data set, and 2) it provides an initial assessment of the impacts of fixed effects (see Table \ref{table:fx} for initial tests of significance). 

The next step is to measure \emph{credibility} by employing \textbf{Criterion 1 (C1)} -- \emph{Markov chains and aKLD or mKLD threshold} -- to test if the responses of identified potential spammers match specific spamming behaviors. When the calculated aKLD or mKLD value is below its threshold, we also propose using a lower-bound threshold on the \textit{average time-to-complete} \footnote{In our case, we use the average time-to-complete as an auxiliary factor and consider accuracy as the ground truth for validation, since true labels are often unavailable in online data collection, which primarily focuses on annotation and labeling tasks. However, when true labels are available, they naturally serve as a more reliable auxiliary criterion. Overall, any additional information, such as time, accuracy, or completeness, beyond the responses themselves can be incorporated as auxiliary factors within our proposed framework.
}
as an auxiliary criterion, similar to the approach in \cite{gadiraju2015understanding}. Workers with average task times lower than the threshold are then subject to \textbf{Criterion 2 (C2)} -- \textit{deletion analysis} -- for further assessment. This is done by deleting candidate spammers one at a time and comparing the model-based deviance distance using a Chi-squared test. Participants who are subject to the \textit{deletion analysis} have been empirically observed to have lower overall accuracy (see Section Experiment), while participants who fail \textbf{Criteria 1 and 2} in addition to the auxiliary criterion are deemed to be spammers. Data obtained from these participants are then excluded from further analysis. 

\begin{figure}[ht]
\caption{2FI interaction plot}
\begin{subfigure}{0.5\textwidth}
  \includegraphics[width=\linewidth]{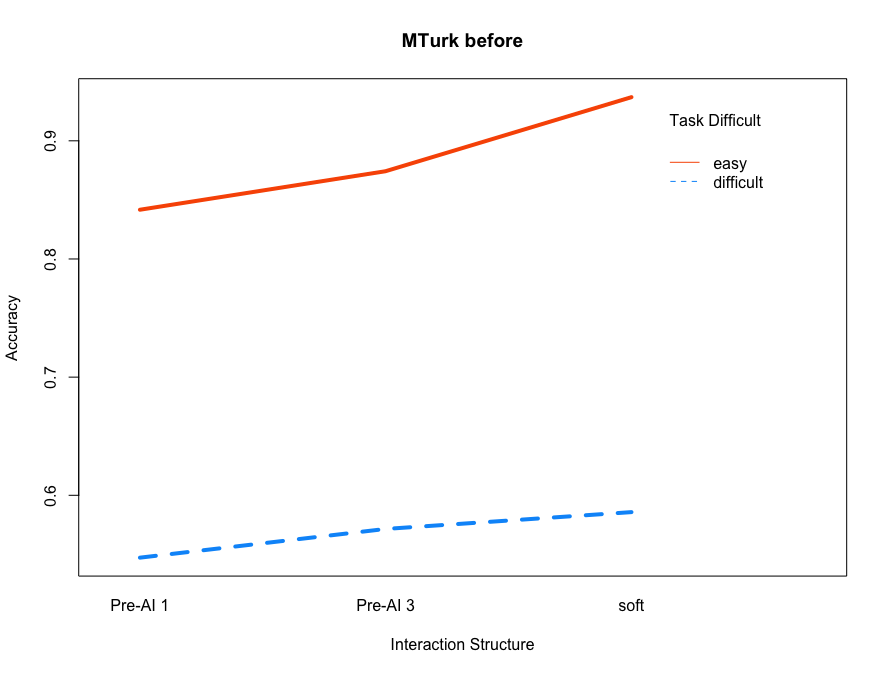}
  \label{fig:2FI before}
\end{subfigure}
\begin{subfigure}{0.5\textwidth}
  \includegraphics[width=\linewidth]{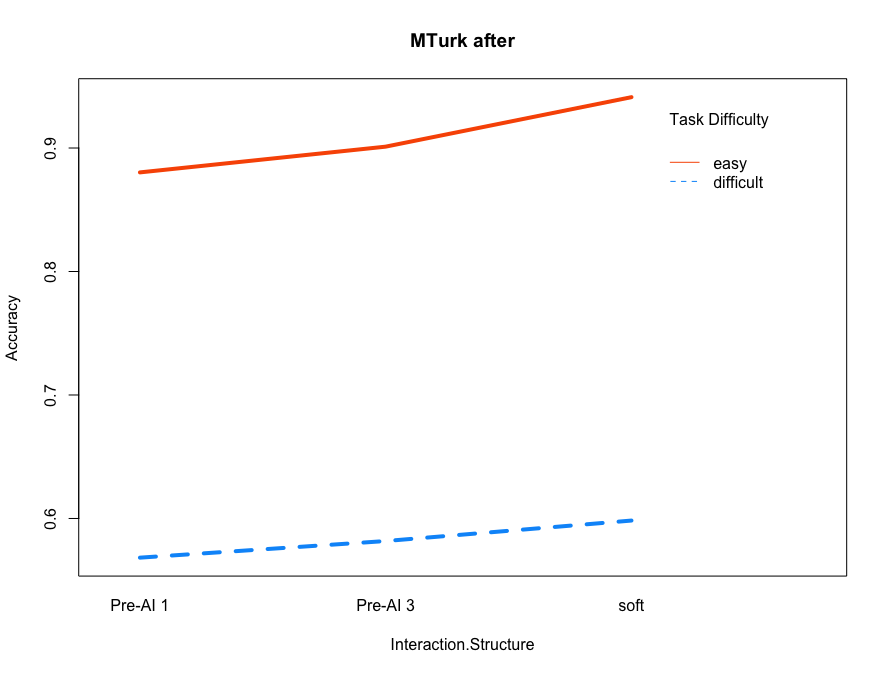}
  \label{fig:2FI after}
\end{subfigure} 
\label{fig:2FI}
\end{figure}

To determine the impact of spammers on fixed effects screening, we proceed to delete their responses from the three datasets and then observe the impact on tests of significance of main and 2FI effects. Table \ref{table:fx} shows the differences in results in Type III Wald $\chi^2$-tests pre- and post-deletion. The reported $p$-values are ANOVA-type tests that are designed for model reduction or variable screening in GLRMs \citep{bolker2015linear}. We find changes in the $p$-values of coefficients for fixed effects in the two crowdsourced datasets: MTurk and Prolific (Table \ref{table:fx}). Specifically, in MTurk, a notable change is observed in the coefficient estimate for the significant two-factor interaction (see Table \ref{table:coef} in Appendix A), resulting in a slight change in the conclusion with regard to the patterns of the impact of the \textit{Interaction Structure} variable (Difficulty: Condition) on the response. Figure \ref{fig:2FI} illustrates that in the post-deletion model, there is congruence between the estimated slopes of \textit{easy} and \textit{difficult} tasks as one changes \textit{Interaction Structure} from one category to another; while in the pre-deletion model, the slopes seem to be different between the two types of tasks. This result highlights the impact of the presence of potential spammers on biases in estimation. 

For the Prolific dataset, deleting all three types of spamming behaviors (Table \ref{table:category}) resulted in changes in the conclusion for the test of significance with respect to the \textit{Condition} variable, with the $p$-value shrinking post-deletion. Potential spamming behavior deletion leads to the reallocation of variance attributed to each predictor, whose changes are captured by the fixed effects tests. On the other hand, no significant changes resulted from deleting possible spamming behaviors for the airport dataset. Note that this experiment was deployed in-person using professionally trained agents in face verification, so it is sensible that due to their professional training and experience, participants' abilities at this task are more consistent with each other i.e., the detected potential spamming behaviors are unlikely to be genuine spammers. Therefore, deleting these observations had little to no impact on the significance tests.

\begin{table}[!ht]
\caption{Comparison of Fixed Effects (Before \& After Identifying \& Removing Potential Spammers)}
\renewcommand{\arraystretch}{1.5}
\centering
\begin{tabular}{@{}*{8}{c}@{}}
\toprule
 Experiment  & Fixed Effects & P-Value Before & P-Value After\\
\midrule
 MTurk     & Intercept    &  0.1214  &   \textbf{0.0379$^\ast$}\\ 
           & Difficulty    &  \textbf{7.312e-16$^\ast$}  &   \textbf{9.563e-16$^\ast$}\\ 
           & Condition    &   0.2715  &   0.7463  \\ \ 
           & Difficulty : Condition   &  \textbf{ 6.588e-11$^\ast$}  &   \textbf{2.357e-05$^\ast$}\\ 
 \midrule
 Prolific  & Intercept    &  0.05110  &   0.05602\\ 
           & Difficulty    & 0.28047   &   0.32726 \\ 
           & Condition    &   0.09019 &   \textbf{0.02082$^\ast$}\\ \ 
           & Difficulty : Condition   &  0.46251  &   0.27657 \\ 

 \midrule
 In Person      & Intercept    &  \textbf{1.255e-05$^\ast$}  &   \textbf{1.258e-05$^\ast$} 
 \\(Airports)   & Difficulty    &  \textbf{3.606e-07$^\ast$}  &   \textbf{4.367e-07$^\ast$}\\ 
                & Condition    &   0.2886   &    0.2073 \\ 
                & Difficulty : Condition   &  0.7168 &   0.8164 \\ 
\bottomrule
\end{tabular}
\footnotetext{$^\ast$ denotes significant in 0.05 critical value} 
\newline 
\newline
\begin{tablenotes} 
\textit{Note:}
Asterisk ($^\ast$) denotes the coefficient estimate significant at a 0.05 critical value; responses of the models are participants' accuracy.
\end{tablenotes}
\label{table:fx}
\end{table}

We recommend that researchers follow the flowchart in Figure \ref{fig:flowchart} as an empirical procedure to assess data quality and identify spamming behaviors. The procedure offers a degree of flexibility with respect to the order of operation and inclusion of either one or both criteria. More specifically, deletion analysis and C2 can come first in the order of steps before C1. In our case, we performed spammer behavior identification first because C1 is more sensitive to potential spammers than C2 and at this point, we do not worry about false detection because there are other criteria (including the auxiliary criterion). Further, if practitioners are not particularly interested in spamming behavior types, then they can go straight to deletion analysis where more severe spamming behaviors can be detected, albeit fewer. However, computational time will increase for a larger pool of workers. Thus, we strongly recommend applying C1 to reduce the pool.

\begin{figure}[p]
\caption{Procedure of Data Quality Evaluation and Spammer Detection}
\resizebox{\textwidth}{!}{%
\begin{tikzpicture}[node distance = 2cm, auto]
    \tikzstyle{decision} = [diamond, draw, 
        text width=8em, 
        text height=0.5em,
        align=center,
        aspect=1.5,
        inner sep=0pt
    ]

    \tikzstyle{block} = [rectangle, draw, 
        text width=8em, text centered, 
        node distance=2cm, 
        minimum height=4em]

    \tikzstyle{input} = [trapezium, draw, 
        trapezium left angle=65, 
        trapezium right angle=115, 
        trapezium stretches,
        text width=7em,
        minimum width=1cm, 
        minimum height=4em, 
        text centered,
        node distance=4cm
    ]

    \tikzstyle{line} = [draw, -latex']
    \tikzstyle{cloud} = [draw, ellipse, node distance=4cm,
        minimum height=4em,
        minimum width=1cm, text centered]

    \node (data) [input] {Crowdsourced data};
    \node (vard) [block, below of=data, yshift=-0.2cm] {Variation decomposition};
    \node (SI)   [decision, below of=vard, yshift=-0.8cm] {Spammer Index > threshold?};
    \node (stop) [cloud, left of=SI, text width=3em, xshift=-0.7cm] {stop};
    \node (mc) [block, below of=SI, yshift=-1cm] {MC and extract transition matrix};
    \node (c1) [decision, below of= mc, yshift=-1cm] {C1: aKLD < empirical thresholds? };
    \node (time1) [decision, right of=c1, xshift=6cm] {is e.g. time, acc < the mean or 1sd?};
    \node (deletion) [block, below of=c1, yshift=-1.5cm] {deletion analysis};
    \node (c2) [decision, below of= deletion, yshift=-1.1cm] {C2: is p-value of deviance distance significant?};
    \node (time2) [decision,right of=c2, xshift=6cm] {is e.g. time, acc < the mean or 1sd?};
    \node (nospammers) [cloud, below of=c2, yshift=0.3cm] {not spammers};
    \node (spammers) [cloud, right of=nospammers, xshift=7.2cm] {spammers};
   
    \path [line] (data) -- (vard);
    \path [line] (vard) -- (SI);
    \path [line] (SI) -- node {Yes} (mc);
    \path [line] (SI) -- node {No}(stop);
    \path [line] (mc) --  (c1);
    \path [line] (c1) -- node [pos=0.25]{No} (deletion);
    \path [line] (c1) -- node [above] {Yes} node [below] {auxiliary factors} (time1);
    \path [line] (time1) |- node [pos=0.2]{No} (0, -13.3);
    \path [line] (deletion) -- (c2);
    \path [line] (c2) -- node [pos=0.3]{No} (nospammers);
    \path [line] (time1) -|   node [pos=0.2]{Yes} (spammers);
    \path [line] (c2) --  node [above] {Yes} node [below] {auxiliary factors}(time2);
    \path [line] (time2) -| node [pos=0.2]{Yes}  (spammers.north);
    \path [line] (time2) |- node [pos=0.2]{No} (0, -20);
\end{tikzpicture}%
}
\label{fig:flowchart}
\end{figure}

\subsection{Limitations}
The metrics and methods proposed in this paper have been shown to be effective at spammer detection, yet they face several limitations that merit attention. The two simulation-based metrics -- aKLD and mKLD -- are robust to unbalanced data (e.g., number of match vs. mismatch cases), which presents an advantage over existing methods. However, establishing thresholds for aKLD or mKLD requires an understanding of task distribution. We suggest using majority voting for positing a preliminary distribution.

Moreover, the variability in datasets necessitates a flexible definition of primary choice spamming behaviors and adjustments in the cutoffs for aKLD or mKLD. This adaptability is guided by theoretical frameworks, specifically the longest run of identical answers theory  \citep{schilling1990longest}, emphasizing the need for a tailored approach to different data scenarios. 

A significant technical challenge is in the precise estimation of the transition probability matrix, a task further exacerbated by the typical volume of tasks assigned to participants. This poses a particular challenge in identifying patterns of \emph{Random Guessing}, a common spamming behavior.

These metrics are proposed with the assumption that the study design closely follows principles for designing statistically valid experiments \citep{montgomery2017design}. However, poorly designed experiments may confound true spamming behavior with artifacts of experimental error as a result of variables not adequately addressed in the experimental design.

In addition, using \emph{time-to-complete} as an auxiliary factor in identifying spammers is not recommended when the imposed time limits are required on task completion, because doing so may confound legitimate behaviors with spamming behaviors. It is recommended that other behavioral metrics based on the time spent on specific components of a task, for instance, to compare time on tasks of difficulty levels rather than on the average time spent, or whether the participant used a specific functionality within the testbed, should instead be explored.

Finally, in our dataset, participant expertise is coupled with the data collection method: non-experts participated online, while experts participated offline. Even though we aim to explore broader applications of our proposed methods to conduct a comparison study in Table \ref{table:category} and a causal inference analysis (Appendix E) to investigate the effects of participants' expertise and data collection methods. Future research could build upon our findings by designing studies that deliberately decouple these variables, for instance, by recruiting online experts and offline non-experts, to allow for a direct and more robust analysis of their independent effects on task performance and behavior.

\section{Conclusion}

In this paper, we propose a framework for determining the quality of crowdsourced data and the identification of possible spamming behaviors in online experiments. Crowdsourced experiments have risen in popularity in the social and behavioral sciences due to their expediency and cost-effectiveness. However, it has been known from previous work that crowdsourced data, particularly from unproctored data collection, are highly susceptible to inconsistent and non-credible responses from participants. Current state-of-the-art methods for evaluating the consistency and credibility of crowdsourced data fall short because of their inability to handle more complex data structures. Furthermore, in many cases, researchers who rely on crowdsourcing platforms are often budget-constrained, making it impossible for them to attempt more trials to collect high-quality data; thus, it is necessary to ensure that responses collected are subject to quality control procedures prior to model building or data analysis. Our major contribution lies in the proposed suite of metrics that can detect potential spammers, followed by identifying posited behavior commonly found among online crowd workers. Through simulations, we show that the Spammer Index serves as an indicator of consistency in data quality and is directly associated with the number of actual spammers artificially imputed into the dataset. Our method is also validated on real-world experimental data that our team collected from Amazon MTurk, Prolific.co, and offline (in-person) data collection with subject matter experts. Through our proposed metrics, the previous finding of better data quality from Prolific.co than Amazon's MTurk is verified. Even though we focus on experiments with binary responses in this paper, our framework is easily extendable to other data scales falling within the exponential family of distributions.


\section*{Declarations}
\subsection{Funding}
This material is based on work supported by the U. S. Department of Homeland Security under Grant Award Number 17STQAC00001-05-00. The views and conclusions contained in this document are those of the authors and should not be interpreted as representing the official policies, either expressed or implied, of the Department of Homeland Security, Department of Defense, or the university.

\subsection{Conflicts of interest/Competing interests}
The authors have no financial or proprietary interests in any material discussed in this article.

\subsection{Ethics approval}
Not applicable

\subsection{Consent to participate}
Informed consent was obtained from all individual participants included in the study.

\subsection{Consent for publication}
Not applicable

\subsection{Availability of data and materials}
 All data simulated or analyzed during this study are included in this published article. 

\subsection{Code availability}
The code was implemented by R. All materials are available in the repository: https://github.com/YangBa78/Data-Reliabilty.

\bibliographystyle{apalike}
\bibliography{refs}

\begin{thebibliography}{}

\bibitem[{Amazon Mechanical Turk}, 2024]{mturk}
{Amazon Mechanical Turk} (2024).
\newblock The online marketplace for work.
\newblock \url{https://www.mturk.com}.

\bibitem[Bates et~al., 2015]{lme4}
Bates, D., M{\"a}chler, M., Bolker, B., and Walker, S. (2015).
\newblock Fitting linear mixed-effects models using {lme4}.
\newblock {\em Journal of Statistical Software}, 67(1):1--48.

\bibitem[Battocchi et~al., 2019]{econml}
Battocchi, K., Dillon, E., Hei, M., Lewis, G., Oka, P., Oprescu, M., and
  Syrgkanis, V. (2019).
\newblock {EconML}: {A Python Package for ML-Based Heterogeneous Treatment
  Effects Estimation}.
\newblock \url{https://github.com/py-why/EconML}.
\newblock Version 0.x.

\bibitem[Belov and Armstrong, 2011]{belov2011distributions}
Belov, D.~I. and Armstrong, R.~D. (2011).
\newblock Distributions of the kullback--leibler divergence with applications.
\newblock {\em British Journal of Mathematical and Statistical Psychology},
  64(2):291--309.

\bibitem[Berger and Zhou, 2014]{berger2014kolmogorov}
Berger, V.~W. and Zhou, Y. (2014).
\newblock Kolmogorov--smirnov test: Overview.
\newblock {\em Wiley Statsref: Statistics Reference Online}.

\bibitem[Bolker, 2015]{bolker2015linear}
Bolker, B.~M. (2015).
\newblock {\em Linear and generalized linear mixed models}, pages 309--333.
\newblock Oxford University Press Oxford, UK.

\bibitem[Browne et~al., 2005]{browne2005variance}
Browne, W.~J., Subramanian, S.~V., Jones, K., and Goldstein, H. (2005).
\newblock Variance partitioning in multilevel logistic models that exhibit
  overdispersion.
\newblock {\em Journal of the Royal Statistical Society Series A: Statistics in
  Society}, 168(3):599--613.

\bibitem[B{\"u}rkner, 2017]{burkner2017brms}
B{\"u}rkner, P.-C. (2017).
\newblock brms: An {R} package for bayesian multilevel models using {S}tan.
\newblock {\em Journal of Statistical Software}, 80:1--28.

\bibitem[Ceschia et~al., 2022]{ceschia2022task}
Ceschia, S., Roitero, K., Demartini, G., Mizzaro, S., Di~Gaspero, L., and
  Schaerf, A. (2022).
\newblock Task design in complex crowdsourcing experiments: Item assignment
  optimization.
\newblock {\em Computers \& Operations Research}, 148:105995.

\bibitem[Chmura~Kraemer et~al., 2002]{chmura2002kappa}
Chmura~Kraemer, H., Periyakoil, V.~S., and Noda, A. (2002).
\newblock Kappa coefficients in medical research.
\newblock {\em Statistics in Medicine}, 21(14):2109--2129.

\bibitem[Christensen, 2023]{ordinal}
Christensen, R. H.~B. (2023).
\newblock {\em ordinal---Regression Models for Ordinal Data}.
\newblock R package version 2023.12-4.1.

\bibitem[Dawid and Skene, 1979]{dawid1979maximum}
Dawid, A.~P. and Skene, A.~M. (1979).
\newblock Maximum likelihood estimation of observer error-rates using the {EM}
  algorithm.
\newblock {\em Journal of the Royal Statistical Society: Series C (Applied
  Statistics)}, 28(1):20--28.

\bibitem[DeMars, 2010]{demars2010item}
DeMars, C. (2010).
\newblock {\em Item Response Theory}.
\newblock Oxford University Press.

\bibitem[Derczynski, 2023]{kappaharm}
Derczynski, L. (2023).
\newblock Kappa scores considered harmful.
\newblock
  \url{https://interhumanagreement.substack.com/p/kappa-scores-considered-harmful}.

\bibitem[Dettori and Norvell, 2020]{dettori2020kappa}
Dettori, J.~R. and Norvell, D.~C. (2020).
\newblock Kappa and beyond: {I}s there agreement?
\newblock {\em Global Spine Journal}, 10(4):499--501.

\bibitem[Difallah et~al., 2018]{difallah2018demographics}
Difallah, D., Filatova, E., and Ipeirotis, P. (2018).
\newblock Demographics and dynamics of mechanical turk workers.
\newblock In {\em Proceedings of the eleventh ACM international conference on
  web search and data mining}, pages 135--143.

\bibitem[Elff, 2024]{elff2022mclogit}
Elff, M. (2024).
\newblock {\em mclogit: Multinomial Logit Models, with or without Random
  Effects or Overdispersion}.
\newblock R package version 0.9.9,.

\bibitem[Estell{\'e}s-Arolas et~al., 2015]{estelles2015crowdsourcing}
Estell{\'e}s-Arolas, E., Navarro-Giner, R., and Gonz{\'a}lez-Ladr{\'o}n-de
  Guevara, F. (2015).
\newblock Crowdsourcing fundamentals: {D}efinition and typology.
\newblock {\em Advances in Crowdsourcing}, pages 33--48.

\bibitem[Falotico and Quatto, 2015]{falotico2015fleiss}
Falotico, R. and Quatto, P. (2015).
\newblock Fleiss' kappa statistic without paradoxes.
\newblock {\em Quality \& Quantity}, 49:463--470.

\bibitem[Feinstein and Cicchetti, 1990]{feinstein1990high}
Feinstein, A.~R. and Cicchetti, D.~V. (1990).
\newblock High agreement but low kappa: I. the problems of two paradoxes.
\newblock {\em Journal of Clinical Epidemiology}, 43(6):543--549.

\bibitem[Fleiss and Cohen, 1973]{fleiss1973equivalence}
Fleiss, J.~L. and Cohen, J. (1973).
\newblock The equivalence of weighted kappa and the intraclass correlation
  coefficient as measures of reliability.
\newblock {\em Educational and Psychological Measurement}, 33(3):613--619.

\bibitem[Gadiraju et~al., 2017]{gadiraju2017using}
Gadiraju, U., Fetahu, B., Kawase, R., Siehndel, P., and Dietze, S. (2017).
\newblock Using worker self-assessments for competence-based pre-selection in
  crowdsourcing microtasks.
\newblock {\em ACM Transactions on Computer-Human Interaction (TOCHI)},
  24(4):1--26.

\bibitem[Gadiraju et~al., 2015]{gadiraju2015understanding}
Gadiraju, U., Kawase, R., Dietze, S., and Demartini, G. (2015).
\newblock Understanding malicious behavior in crowdsourcing platforms: The case
  of online surveys.
\newblock In {\em Proceedings of the 33rd Annual ACM Conference on Human
  Factors in Computing Systems}, pages 1631--1640.

\bibitem[Gamer et~al., 2012]{gamer2012package}
Gamer, M., Lemon, J., Fellows, I., and Singh, P. (2012).
\newblock Package 'irr'. various coefficients of interrater reliability and
  agreement.
\newblock {\em Version 0.84. http://CRAN. R-project. org/package= irr}.

\bibitem[GM et~al., 2021]{gm2021urban}
GM, V., Pereira, B., and Little, S. (2021).
\newblock Urban footpath image dataset to assess pedestrian mobility.
\newblock In {\em Proceedings of the 1st International Workshop on Multimedia
  Computing for Urban Data}, pages 23--30.

\bibitem[Grilli and Rampichini, 2007]{grilli2007multilevel}
Grilli, L. and Rampichini, C. (2007).
\newblock A multilevel multinomial logit model for the analysis of graduates'
  skills.
\newblock {\em Statistical Methods and Applications}, 16:381--393.

\bibitem[Hall et~al., 2022]{hall2022quality}
Hall, M., Afzali, M.~F., Krause, M., and Caton, S. (2022).
\newblock What quality control mechanisms do we need for high-quality crowd
  work?
\newblock {\em IEEE Access}, 10:99709--99723.

\bibitem[Hovy et~al., 2013]{hovy2013learning}
Hovy, D., Berg-Kirkpatrick, T., Vaswani, A., and Hovy, E. (2013).
\newblock Learning whom to trust with {MACE}.
\newblock In Vanderwende, L., Daum{\'e}~III, H., and Kirchhoff, K., editors,
  {\em Proceedings of the 2013 Conference of the North {A}merican Chapter of
  the Association for Computational Linguistics: Human Language Technologies},
  pages 1120--1130, Atlanta, Georgia. Association for Computational
  Linguistics.

\bibitem[Hsueh et~al., 2009]{hsueh2009data}
Hsueh, P.-Y., Melville, P., and Sindhwani, V. (2009).
\newblock Data quality from crowdsourcing: A study of annotation selection
  criteria.
\newblock In {\em Proceedings of the NAACL HLT 2009 Workshop on Active Learning
  for Natural Language Processing}, pages 27--35.

\bibitem[Huff and Tingley, 2015]{huff2015these}
Huff, C. and Tingley, D. (2015).
\newblock "{W}ho are these people?" {E}valuating the demographic
  characteristics and political preferences of {MT}urk survey respondents.
\newblock {\em Research \& Politics}, 2(3):2053168015604648.

\bibitem[Ipeirotis et~al., 2014]{ipeirotis2014repeated}
Ipeirotis, P.~G., Provost, F., Sheng, V.~S., and Wang, J. (2014).
\newblock Repeated labeling using multiple noisy labelers.
\newblock {\em Data Mining and Knowledge Discovery}, 28:402--441.

\bibitem[Junkes et~al., 2015]{junkes2015validity}
Junkes, M.~C., Fraiz, F.~C., Sardenberg, F., Lee, J.~Y., Paiva, S.~M., and
  Ferreira, F.~M. (2015).
\newblock Validity and reliability of the brazilian version of the rapid
  estimate of adult literacy in dentistry--breald-30.
\newblock {\em PloS One}, 10(7):e0131600.

\bibitem[Landis and Koch, 1977]{landis1977measurement}
Landis, J.~R. and Koch, G.~G. (1977).
\newblock The measurement of observer agreement for categorical data.
\newblock {\em Biometrics}, pages 159--174.

\bibitem[Li et~al., 2016]{li2016noise}
Li, C., Sheng, V.~S., Jiang, L., and Li, H. (2016).
\newblock Noise filtering to improve data and model quality for crowdsourcing.
\newblock {\em Knowledge-Based Systems}, 107:96--103.

\bibitem[Liang et~al., 2018]{liang2018intrinsic}
Liang, H., Wang, M.-M., Wang, J.-J., and Xue, Y. (2018).
\newblock How intrinsic motivation and extrinsic incentives affect task effort
  in crowdsourcing contests: A mediated moderation model.
\newblock {\em Computers in Human Behavior}, 81:168--176.

\bibitem[Liljequist et~al., 2019]{liljequist2019intraclass}
Liljequist, D., Elfving, B., and Skavberg~Roaldsen, K. (2019).
\newblock Intraclass correlation--a discussion and demonstration of basic
  features.
\newblock {\em PloS one}, 14(7):e0219854.

\bibitem[McHugh, 2012]{mchugh2012interrater}
McHugh, M.~L. (2012).
\newblock Interrater reliability: {T}he kappa statistic.
\newblock {\em Biochemia Medica}, 22(3):276--282.

\bibitem[Mehta et~al., 2018]{mehta2018performance}
Mehta, S., Bastero-Caballero, R.~F., Sun, Y., Zhu, R., Murphy, D.~K., Hardas,
  B., and Koch, G. (2018).
\newblock Performance of intraclass correlation coefficient ({ICC}) as a
  reliability index under various distributions in scale reliability studies.
\newblock {\em Statistics in Medicine}, 37(18):2734--2752.

\bibitem[Meyer et~al., 2016]{meyer2016net}
Meyer, E.~T., Schroeder, R., and Cowls, J. (2016).
\newblock The net as a knowledge machine: How the internet became embedded in
  research.
\newblock {\em New Media \& Society}, 18(7):1159--1189.

\bibitem[Montgomery, 2009]{montgomery2009statistical}
Montgomery, D.~C. (2009).
\newblock {\em Statistical Quality Control}, volume~7.
\newblock Wiley New York.

\bibitem[Montgomery, 2017]{montgomery2017design}
Montgomery, D.~C. (2017).
\newblock {\em Design and Analysis of Experiments}.
\newblock John Wiley \& Sons.

\bibitem[Peer et~al., 2022]{peer2022data}
Peer, E., Rothschild, D., Gordon, A., Evernden, Z., and Damer, E. (2022).
\newblock Data quality of platforms and panels for online behavioral research.
\newblock {\em Behavior Research Methods}, 54:1643--1662.

\bibitem[Peng et~al., 2013]{peng2013mapping}
Peng, T.-Q., Zhang, L., Zhong, Z.-J., and Zhu, J.~J. (2013).
\newblock Mapping the landscape of internet studies: Text mining of social
  science journal articles 2000--2009.
\newblock {\em {N}ew {M}edia {\&} {S}ociety}, 15(5):644--664.

\bibitem[P{\'e}rez-Rosas et~al., 2016]{perez2016building}
P{\'e}rez-Rosas, V., Mihalcea, R., Resnicow, K., Singh, S., and An, L. (2016).
\newblock Building a motivational interviewing dataset.
\newblock In {\em Proceedings of the Third Workshop on Computational
  Linguistics and Clinical Psychology}, pages 42--51.

\bibitem[Pierce and Schafer, 1986]{pierce1986residuals}
Pierce, D.~A. and Schafer, D.~W. (1986).
\newblock Residuals in generalized linear models.
\newblock {\em Journal of the American Statistical Association},
  81(396):977--986.

\bibitem[{Prolific}, 2024]{prolific}
{Prolific} (2024).
\newblock Quickly find research participants you can trust.
\newblock \url{https://www.prolific.com}.

\bibitem[Riezler and Hagmann, 2022]{riezler2022validity}
Riezler, S. and Hagmann, M. (2022).
\newblock {\em Validity, Reliability, and Significance: Empirical Methods for
  NLP and Data Science}.
\newblock Springer Nature.

\bibitem[Rodriguez and Oppenheimer, 2024]{rodriguez2023creating}
Rodriguez, C. and Oppenheimer, D.~M. (2024).
\newblock Creating a bot-tleneck for malicious {AI}: Psychological methods for
  bot detection.
\newblock {\em Behavior Research Methods}, pages 1--18.

\bibitem[Salehi et~al., 2023]{salehi2023evaluating}
Salehi, P., Ba, Y., Kim, N., Mosallanezhad, A., Pan, A., Cohen, M.~C., Wang,
  Y., Zhao, J., Bhatti, S., Sung, J., et~al. (2023).
\newblock Evaluating trustworthiness of {AI-Enabled} decision support systems:
  Validation of the multisource {AI} scorecard table ({MAST}).
\newblock {\em arXiv preprint arXiv:2311.18040}.

\bibitem[Salminen et~al., 2018]{salminen2018inter}
Salminen, J.~O., Al-Merekhi, H.~A., Dey, P., and Jansen, B.~J. (2018).
\newblock Inter-rater agreement for social computing studies.
\newblock In {\em 2018 Fifth International Conference on Social Networks
  Analysis, Management and Security (SNAMS)}, pages 80--87.

\bibitem[Schilling, 1990]{schilling1990longest}
Schilling, M.~F. (1990).
\newblock The longest run of heads.
\newblock {\em The College Mathematics Journal}, 21(3):196--207.

\bibitem[Sheng and Zhang, 2019]{sheng2019machine}
Sheng, V.~S. and Zhang, J. (2019).
\newblock Machine learning with crowdsourcing: A brief summary of the past
  research and future directions.
\newblock In {\em Proceedings of the AAAI Conference on Artificial
  Intelligence}, volume~33, pages 9837--9843.

\bibitem[Smyth, 2003]{smyth2003pearson}
Smyth, G.~K. (2003).
\newblock Pearson's goodness of fit statistic as a score test statistic.
\newblock {\em Lecture Notes-Monograph Series}, 40:115--126.

\bibitem[Spedicato, 2017]{spedicato2017discrete}
Spedicato, G.~A. (2017).
\newblock Discrete time markov chains with r.
\newblock {\em The R Journal}, 9(2):84.

\bibitem[Steger et~al., 2018]{steger2018meta}
Steger, D., Schroeders, U., and Gnambs, T. (2018).
\newblock A meta-analysis of test scores in proctored and unproctored ability
  assessments.
\newblock {\em European Journal of Psychological Assessment}.

\bibitem[Tao et~al., 2018]{tao2018domain}
Tao, D., Cheng, J., Yu, Z., Yue, K., and Wang, L. (2018).
\newblock Domain-weighted majority voting for crowdsourcing.
\newblock {\em IEEE Transactions on Neural Networks and Learning Systems},
  30(1):163--174.

\bibitem[Traub, 1997]{traub1997classical}
Traub, R.~E. (1997).
\newblock Classical test theory in historical perspective.
\newblock {\em Educational Measurement}, 16:8--13.

\bibitem[Van~Exel et~al., 2010]{van2010impact}
Van~Exel, M., Dias, E., and Fruijtier, S. (2010).
\newblock The impact of crowdsourcing on spatial data quality indicators.
\newblock In {\em Proceedings of the GIScience 2010 Doctoral Colloquium,
  Zurich, Switzerland}, pages 14--17.

\bibitem[Wang et~al., 2020]{wang2020collueagle}
Wang, Z., Hu, R., Chen, Q., Gao, P., and Xu, X. (2020).
\newblock Collueagle: collusive review spammer detection using markov random
  fields.
\newblock {\em Data Mining and Knowledge Discovery}, 34:1621--1641.

\bibitem[Waseem, 2016]{waseem2016you}
Waseem, Z. (2016).
\newblock Are you a racist or am {I} seeing things? annotator influence on hate
  speech detection on {T}witter.
\newblock In Bamman, D., Do{\u{g}}ru{\"o}z, A.~S., Eisenstein, J., Hovy, D.,
  Jurgens, D., O{'}Connor, B., Oh, A., Tsur, O., and Volkova, S., editors, {\em
  Proceedings of the First Workshop on {NLP} and Computational Social Science},
  pages 138--142, Austin, Texas. Association for Computational Linguistics.

\bibitem[Webb et~al., 2006]{webb20064}
Webb, N.~M., Shavelson, R.~J., and Haertel, E.~H. (2006).
\newblock 4 reliability coefficients and generalizability theory.
\newblock {\em Handbook of statistics}, 26:81--124.

\bibitem[Wilks, 1938]{wilks1938large}
Wilks, S.~S. (1938).
\newblock The large-sample distribution of the likelihood ratio for testing
  composite hypotheses.
\newblock {\em The Annals of Mathematical Statistics}, 9(1):60--62.

\bibitem[Zhang et~al., 2018]{zhang2018understanding}
Zhang, Y., Ding, X., and Gu, N. (2018).
\newblock Understanding fatigue and its impact in crowdsourcing.
\newblock In {\em 2018 IEEE 22nd International Conference on Computer Supported
  Cooperative Work in Design ((CSCWD))}, pages 57--62.

\bibitem[Zhou and Fishbach, 2016]{zhou2016pitfall}
Zhou, H. and Fishbach, A. (2016).
\newblock The pitfall of experimenting on the web: How unattended selective
  attrition leads to surprising (yet false) research conclusions.
\newblock {\em Journal of Personality and Social Psychology}, 111(4):493.

\end{thebibliography}
\clearpage
\section{Appendix A}

\begin{table}[ht]
\caption{Comparison of Fixed Effects Coefficients (Before \& After Identifying \& Removing Potential Spammers)}
\renewcommand{\arraystretch}{1.5}
\begin{tabular}{@{}*{8}{c}@{}}
\toprule
 Experiment  & Fixed Effects & Coefficients Before & Coefficients After\\
\midrule
 MTurk     & Intercept    &  0.24218  &   \textbf{0.36485}\\ 
           & Task.Easy    &  \textbf{1.72905}  &   \textbf{1.98470}\\ 
           & Condition.Pre-AI3    &   0.07139  &   0.04418\\ 
           & Condition.Soft.Deferral   &  0.12781   &   0.06879\\ 
           & Task.Easy: Condition.Pre-AI3    &  0.17292   &   0.18483\\ 
           & Task.Easy: Condition.Soft.Deferral   &  \textbf{0.81746}  &   \textbf{0.65937}\\ 
 \midrule
 Prolific  & Intercept    &  -1.0411  &   -1.0779  \\ 
           & Task.Easy    &  0.8197   &   0.7888\\ 
           & Condition.Soft.Deferral    &  0.2113  &   \textbf{0.2615}\\ 
           & Task.Easy: Condition.Soft.Deferral    &   0.1511  &   0.2447 \\ 

 \midrule
 In Person     & Intercept    &  \textbf{0.89985}  &   \textbf{0.94458}
 \\(Airports)    & Task.Easy    &  \textbf{1.44520}  &   \textbf{1.51685} \\ 
                  & Condition.LowMAST     &  -0.10571  &   -0.12799  \\ 
                  & Task.Easy: Condition.LowMAST    &  -0.03637  &   0.02512 \\ 
\bottomrule
\end{tabular}
\newline 
\newline
\begin{tablenotes} 
\textit{Note:}
The bold number denotes the coefficient estimate significant in 0.05 critical value; responses of the models are participants' accuracy.
\end{tablenotes}
\label{table:coef}
\end{table}

\clearpage
\section{Appendix B}
\label{appendix:b}

This Appendix describes the simulation-based study for ordinal and nominal response data.

\noindent \textbf{Simulations strategy in ordinal response data.}
Scale Likert is predefined as 1-5 (1 < 2 < 3 < 4 < 5). Each task is assigned a latent continuous value (random effects, $image_{eff}$) drawn from a normal distribution with variance 6. The random effects values are divided into ordered categories using quantities as cutoffs. 
\begin{description}
    \item[Normal workers:] Random effects for workers ($worker_{eff}$) and their interactions ($int_{eff}$) are randomly sampled from [-0.4, 0.4]. The normal worker's final score is the summation of all random effects ($image_{eff} + worker_{eff} + int_{eff}$), and their decision is categorized using the same quantile-based cutoffs as the true labels. 
    \item[Random Guessing Spammers:] Each worker's decision is sampled from the five ordinal categories with equal probability (20\% for each class).
    \item[Primary Choice Spammers:] Workers heavily favor a single class with an 88\% probability, while all other classes share the remaining probability.
    \item[Strong Pattern Spammers:] Workers follow a systematic cycle in their responses through $(i \, \%\% \, n_{\text{classes}}) + 1$ with 96\% probability. Below is the transition probability matrix for the simulated samples of the three types of spammers.
\end{description} 

\[\resizebox{\textwidth}{!}{$
\begin{array}{ccc}
    \mathbf{P}_{\text{workerID=4}} = 
    \begin{blockarray}{cccccc}
        & 1    & 2    & 3    & 4    & 5  \\
        \begin{block}{r(rrrrr)}
        1 & 0.32 & 0.18 & 0.18 & 0.09 & 0.23 \\
        2 & 0.25 & 0.12 & 0.00 & 0.12 & 0.50 \\
        3 & 0.20 & 0.30 & 0.00 & 0.30 & 0.20 \\
        4 & 0.27 & 0.27 & 0.18 & 0.18 & 0.09 \\
        5 & 0.30 & 0.15 & 0.25 & 0.10 & 0.20 \\
        \end{block}
    \end{blockarray} &
    \mathbf{P}_{\text{workerID=7}} = 
    \begin{blockarray}{cccccc}
        & 1    & 2    & 3    & 4    & 5  \\
        \begin{block}{r(rrrrr)}
        1 & 0.00 & 0.00 & 1.00 & 0.00 & 0.00 \\
        2 & 0.00 & 0.00 & 1.00 & 0.00 & 0.00 \\
        3 & 0.01 & 0.03 & 0.90 & 0.03 & 0.03 \\
        4 & 0.00 & 0.00 & 1.00 & 0.00 & 0.00 \\
        5 & 0.00 & 0.00 & 1.00 & 0.00 & 0.00 \\
        \end{block}
    \end{blockarray} &
    \mathbf{P}_{\text{workerID=13}} = 
    \begin{blockarray}{cccccc}
        & 1    & 2    & 3    & 4    & 5  \\
        \begin{block}{r(rrrrr)}
        1 & 0.00 & 1.00 & 0.00 & 0.00 & 0.00 \\
        2 & 0.00 & 0.06 & 0.89 & 0.00 & 0.06 \\
        3 & 0.00 & 0.06 & 0.00 & 0.94 & 0.00 \\
        4 & 0.00 & 0.00 & 0.00 & 0.00 & 1.00 \\
        5 & 0.82 & 0.12 & 0.00 & 0.00 & 0.06 \\
        \end{block}
    \end{blockarray}
\end{array}
$}\]

\noindent \textbf{Simulations strategy in nominal response data.} The category is predefined as "A", "B", "C", "D" and "E". Most procedures remain the same for simulating ordinal data, except that five categories are assigned using quantities as cutoffs based on the random effects of images, after which the categories are randomized to generate the final labels. Below is the transition probability matrix for the simulated samples of the three types of spammers.

\[
\resizebox{\textwidth}{!}{$
\begin{array}{ccc}
    \mathbf{P}_{\text{workerID=1}} = 
    \begin{blockarray}{cccccc}
        & A & B & C & D & E \\
        \begin{block}{r(rrrrr)}
        A & 0.27 & 0.13 & 0.20 & 0.20 & 0.20 \\
        B & 0.17 & 0.25 & 0.25 & 0.25 & 0.08 \\
        C & 0.17 & 0.06 & 0.22 & 0.33 & 0.22 \\
        D & 0.20 & 0.25 & 0.20 & 0.20 & 0.15 \\
        E & 0.14 & 0.07 & 0.29 & 0.29 & 0.21 \\
        \end{block}
    \end{blockarray} &

    \mathbf{P}_{\text{workerID=9}} = 
    \begin{blockarray}{cccccc}
        & A & B & C & D & E \\
        \begin{block}{r(rrrrr)}
        A & 0.00 & 0.00 & 0.00 & 0.00 & 1.00 \\
        B & 0.00 & 0.00 & 0.00 & 0.00 & 1.00 \\
        C & 0.00 & 0.00 & 0.00 & 0.00 & 1.00 \\
        D & 0.14 & 0.00 & 0.14 & 0.00 & 0.71 \\
        E & 0.03 & 0.02 & 0.06 & 0.11 & 0.78 \\
        \end{block}
    \end{blockarray} &

    \mathbf{P}_{\text{workerID=13}} = 
    \begin{blockarray}{cccccc}
        & A & B & C & D & E \\
        \begin{block}{r(rrrrr)}
        A & 0.06 & 0.94 & 0.00 & 0.00 & 0.00 \\
        B & 0.00 & 0.00 & 1.00 & 0.00 & 0.00 \\
        C & 0.00 & 0.00 & 0.00 & 1.00 & 0.00 \\
        D & 0.06 & 0.00 & 0.00 & 0.00 & 0.94 \\
        E & 1.00 & 0.00 & 0.00 & 0.00 & 0.00 \\
        \end{block}
    \end{blockarray}
\end{array}
$}
\]

\clearpage
\section{Appendix C}

\label{appendix:c}

\begin{figure}[ht]
\caption{Halfway Spamming Behavior Detection}
\begin{subfigure}{0.5\textwidth}
  \includegraphics[width=\linewidth]{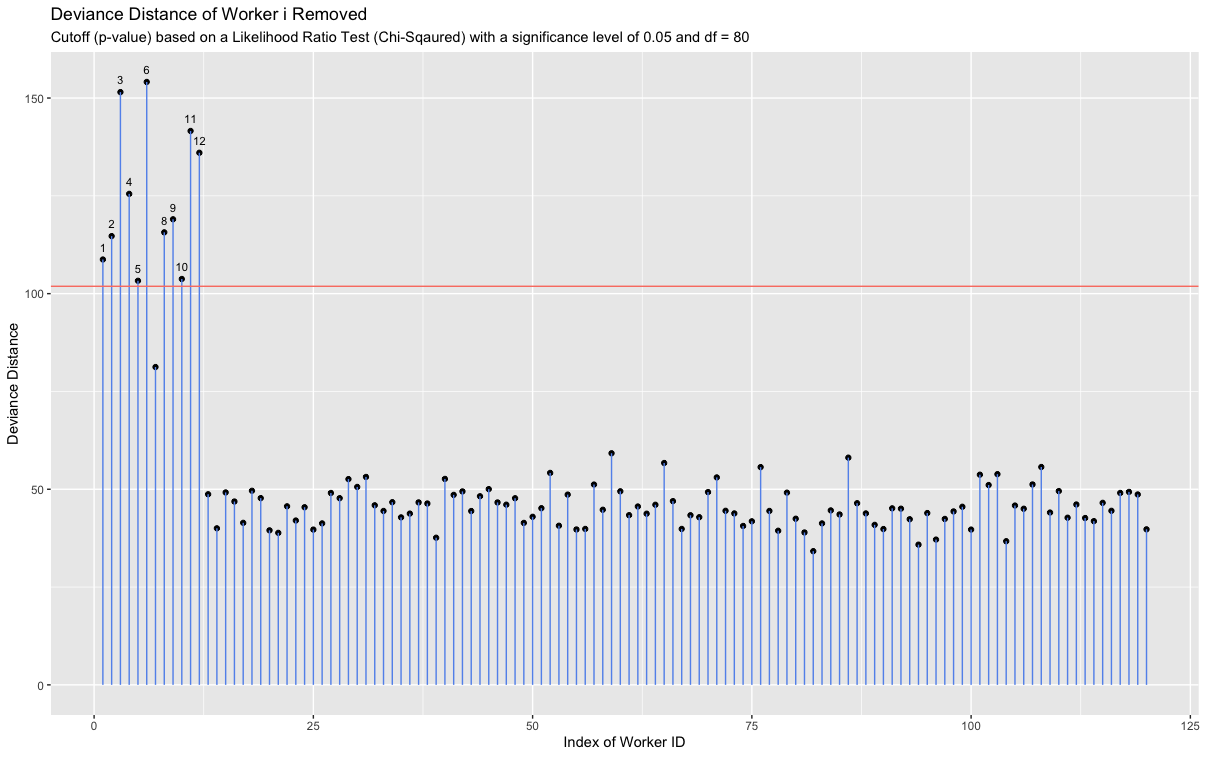}
  \caption{Deviance distance for simulated halfway spamming behaviors}
  \label{fig:half-1}
\end{subfigure}
\begin{subfigure}{0.5\textwidth}
  \includegraphics[width=\linewidth]{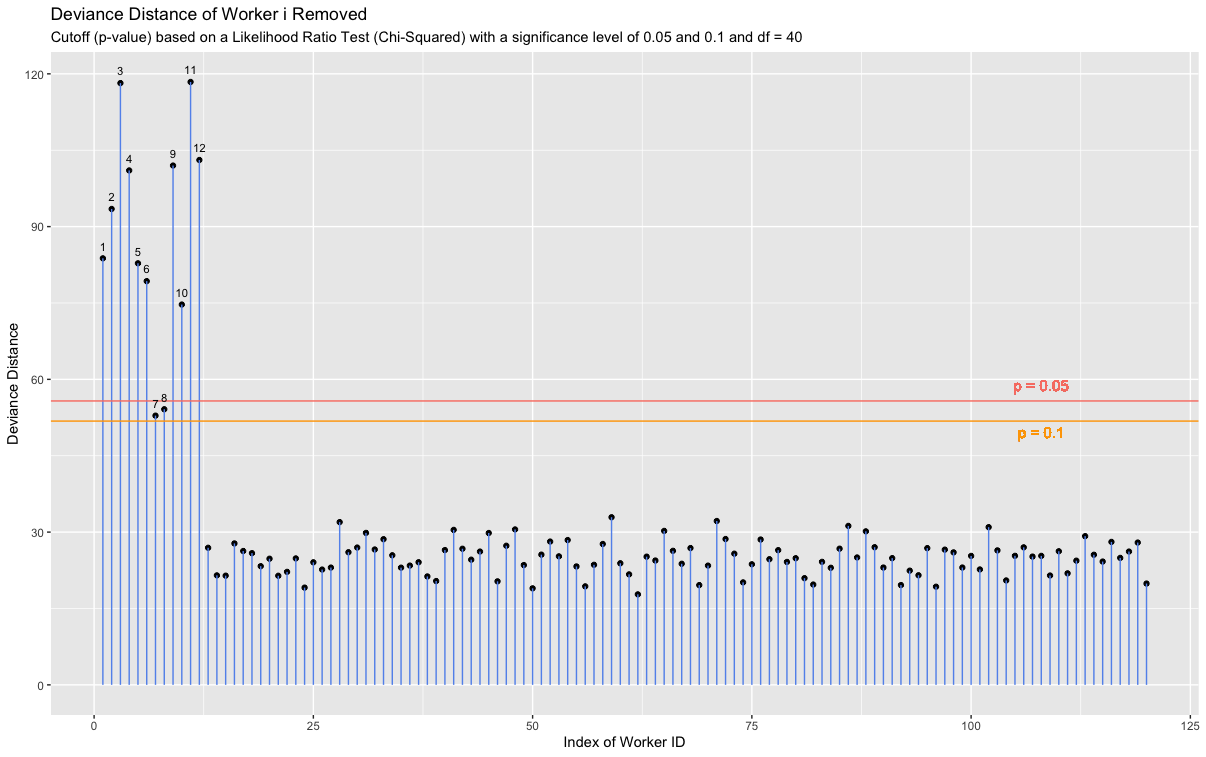}
  \caption{Deviance distances for a subset (Tasks 41 - task 80)}
  \label{fig:half-half}
\end{subfigure} 
\label{fig:half}
\end{figure}

Following the procedure outlined in the Simulation Section, we simulated 12 spammers (Worker ID 1-12) and replaced their first 40 tasks with the decisions of normal workers, introducing their spamming behaviors starting from task 41 to task 80. Figure \ref{fig:half} presents the results of the deletion analysis. By applying this analysis to the entire dataset, we were able to identify 11 out of 12 spammers using a threshold of p = 0.05. (Figure \ref{fig:half-1}). When performing the deletion analysis on the subset of tasks 41 through 80, with a p-value threshold of 0.1, all spammers are detected with 40 degrees of freedom in a likelihood-ratio test(Figure \ref{fig:half-half}). In practice, deletion analysis can be combined with the sliding window approach to identify spamming behaviors within fixed task intervals. Additionally, the p-value obtained from the likelihood-ratio test can serve as a flexible threshold.
\newline

\clearpage
\section{Appendix D}

\label{appendix:d}

\clearpage
\section{Appendix E}
\label{appendix:e}

\begin{figure}[ht]
\caption{graphs of causal models}
\begin{subfigure}{0.45\textwidth}
  \includegraphics[width=\linewidth]{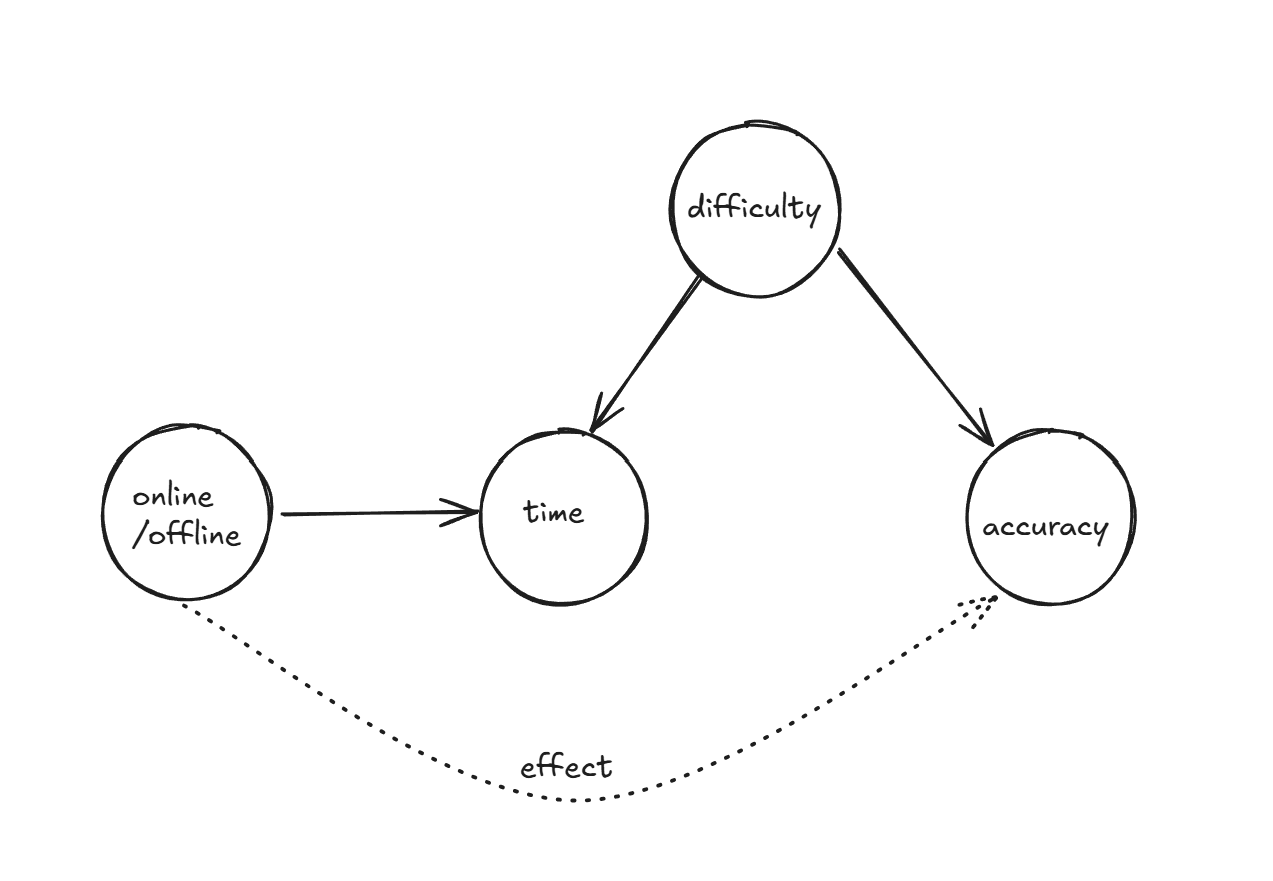}
  \caption{model 1}
  \label{fig:m1}
\end{subfigure}
\begin{subfigure}{0.42\textwidth}
  \includegraphics[width=\linewidth]{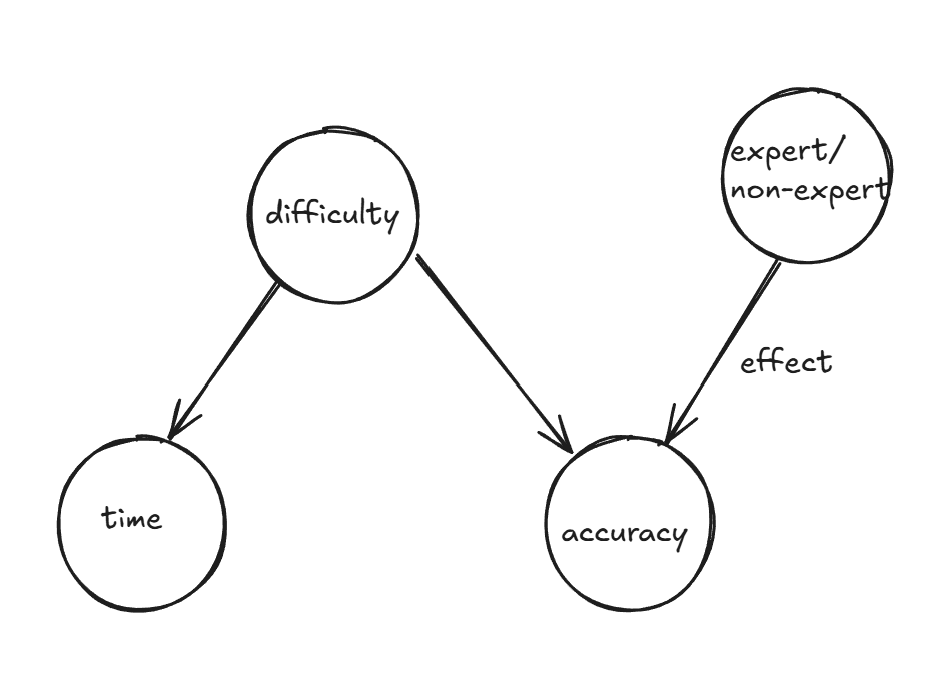}
  \caption{model 2}
  \label{fig:m2}
\end{subfigure} 
\label{fig:causal models}
\end{figure}

In Table \ref{table:category}, we validate the effectiveness of our approach through three different datasets, including online and offline collections. Our objective is to provide broader applications of our proposed methods rather than focusing solely on online and offline differences. To further investigate whether the observed differences are mainly due to participants' skills or the data collection method, we leverage causal inference to analyze our data. Figure \ref{fig:causal models} demonstrates our two hypothesized causal models. In Model 1, we assume that the mode of data collection (online vs. offline) influences the time spent on each task since online participants may aim to maximize their earnings rate, while task difficulty affects both time and accuracy. However, we do not expect the mode of data collection to have a direct effect on accuracy. In contrast, Model 2 maintains the same assumptions regarding task difficulty, impacting both time and accuracy, but additionally posits that participants' expertise directly influences accuracy.

\begin{table}[ht]
\centering
\caption{Causal Models Result}
\begin{tabular}{lrrrrr}
\toprule
 & Estimate & StdErr & P-Value & CI Lower & CI Upper \\
\midrule
 \textbf{Model 1} & -0.070259 & 0.004989 & $4.82 \times 10^{-45}$ & -0.080037 & -0.060481 \\
 \midrule
 \textbf{Model 2} & -0.124093 & 0.006335 & $1.91 \times 10^{-85}$ & -0.136509 & -0.111678 \\
\bottomrule
\end{tabular}
\label{tab:causal}
\end{table}

We use EconML \citep{econml} package in Python to implement both models, and Table \ref{tab:causal} is the result. It shows that switching from offline to online leads to a 7\% decrease in expected accuracy, while non-experts perform with 12\% lower accuracy compared to experts. Also, Model 2 yields a significantly smaller p-value than Model 1, indicating a better fit to the data by Model 2. Therefore, we conclude that the observed differences in task performance (accuracy) between the datasets are primarily driven by participants' expertise.


\end{document}